%% file: main.tex
\PassOptionsToPackage{table}{xcolor}
\documentclass[sigconf,nonacm]{acmart}
\AtBeginDocument{%
  \providecommand\BibTeX{{%
    \normalfont B\kern-0.5em{\scshape i\kern-0.25em b}\kern-0.8em\TeX}}}

\usepackage{amsmath,amsfonts}
\usepackage{algorithmic}
\usepackage{graphicx}
\usepackage{xcolor}
\usepackage{xspace}
\usepackage{xurl}
\usepackage{hyperref}

\usepackage{listings}
\newcommand{\codeinline}[1]{\lstinline[basicstyle=\ttfamily,breaklines=true,breakatwhitespace=true]{#1}}
\usepackage{enumitem}
\usepackage{multirow}
\usepackage{booktabs}
\usepackage{array}
\usepackage{subcaption}
\usepackage{pifont}
\usepackage{fancyvrb}
\usepackage{optidef}

\usepackage{forest}
\usepackage{fontawesome} 

\usepackage[english]{babel}

\usepackage{tikz}
\usetikzlibrary{shapes.geometric, arrows}

\usepackage{multicol}
\usepackage{adjustbox}
\usepackage{makecell}
\usepackage{colortbl}
\usepackage[type={CC},
modifier={by-sa},
version={4.0}]{doclicense}

\usepackage[colorinlistoftodos]{todonotes}

\usepackage[ruled,linesnumbered]{algorithm2e}

\newcommand{\cmark}{\ding{51}}%
\newcommand{\xmark}{\ding{55}}%

\usepackage[acronym]{glossaries}

\addto\extrasenglish{
  
}

\addto\extrasenglish{
  
}

\graphicspath{{figs/}}

\hypersetup{%
  colorlinks=true,
  linkcolor=black,
  citecolor=black,
  urlcolor=black
}

\newacronym{ML}{ML}{Machine Learning}
\newacronym{AI}{AI}{Artificial Intelligence}
\newacronym{TNR}{TNR}{True Negative Rate}
\newacronym{FPR}{FPR}{False Positive Rate}
\newacronym{TPR}{TPR}{True Positive Rate}
\newacronym{ROC}{ROC}{Receiver Operating Characteristic}
\newacronym{DR}{DR}{Detection Rate}
\newacronym{IOC}{IoC}{Indicators of Compromise}

\graphicspath{{figures/}}

\newcolumntype{C}[1]{>{\centering\let\newline\\\arraybackslash\hspace{0pt}}m{#1}}

\makeatletter
\AtBeginDocument{\let\c@listing\c@lstlisting}
\makeatother

\begin{document}

\title[The Role of Domain-Specific Features in Malware Detection: A macOS Case Study]{The Role of Domain-Specific Features in Malware Detection: \\ A macOS Case Study}

\author{Biagio Montaruli}
\email{montarul@eurecom.fr}
\affiliation{%
  \institution{EURECOM \& SAP}
  \country{France}
}

\author{Andrea Oliveri}
\email{oliveri@eurecom.fr}
\affiliation{%
  \institution{EURECOM}
  \country{France}
}

\author{Savino Dambra}
\email{savino.dambra@gendigital.com}
\affiliation{%
  \institution{GenDigital}
  \country{France}
}

\author{Davide Balzarotti}
\email{balzarot@eurecom.fr}
\affiliation{%
  \institution{EURECOM}
  \country{France}
}

\renewcommand{\shortauthors}{Montaruli, et al.}

\input{commands}

\input{src/abstract}



\acmYear{2026}\copyrightyear{2026}
\setcopyright{cc}
\setcctype[4.0]{by}
\acmConference[ASIA CCS '26]{ACM Asia Conference on Computer and Communications Security}{June 1--5, 2026}{Bangalore, India}
\acmBooktitle{ACM Asia Conference on Computer and Communications Security (ASIA CCS '26), June 1--5, 2026, Bangalore, India}
\acmDOI{10.1145/3779208.3785392}
\acmISBN{979-8-4007-2356-8/26/06}

\maketitle

\input{src/1-introduction}
\input{src/2-background}
\input{src/3-features}
\input{src/4-dataset}
\input{src/5-experiments}
\input{src/6-related}

\input{src/7-conclusions}

\bibliographystyle{ACM-Reference-Format}
\bibliography{bibliography}

\appendix
\input{src/appendix}
\end{document}

%% file: commands.tex
\newcommand{\sota}{state-of-the-art\xspace}
\newcommand{\diff}[2]{\frac{\partial #1}{\partial #2}}
\newcommand{\vct}[1]{\ensuremath{\boldsymbol{#1}}}
\newcommand{\mat}[1]{\ensuremath{\mathbf{#1}}}
\newcommand{\set}[1]{\ensuremath{\mathcal{#1}}}
\newcommand{\con}[1]{\ensuremath{\mathsf{#1}}}
\newcommand{\tsum}{\ensuremath{\textstyle \sum}}
\newcommand{\T}{\ensuremath{\top}}
\newcommand{\mycomment}[1]{\footnote{\textcolor{red}{#1}}}
\newcommand{\ind}[1]{\ensuremath{\mathbbm 1_{#1}}}
\newcommand{\argmax}{\operatornamewithlimits{\arg\,\max}}
\newcommand{\erf}{\text{erf}}
\newcommand{\sign}{\text{sign}}
\newcommand{\argmin}{\operatornamewithlimits{\arg\,\min}}
\newcommand{\bmat}[1]{\begin{bmatrix}#1\end{bmatrix}}
\newcommand{\myparagraph}[1]{\noindent \textbf{#1}}
\newcommand{\myparagraphlb}[1]{\noindent \newline \textbf{#1}}
\newcommand{\mysubparagraph}[1]{\noindent \underline{\textit{#1}}}
\newcommand{\ie}{i.e.\xspace}
\newcommand{\eg}{e.g.\xspace}
\newcommand{\etc}{etc.\xspace}
\newcommand{\aka}{a.k.a.\xspace}
\newcommand{\wrt}{w.r.t.\xspace}
\newcommand{\etal}{\emph{et al.}\xspace}
\newcommand{\macho}{Mach-O\xspace}

\newcommand{\folderapp}{\faFolder}
\newcommand{\fileapp}{\faFile}
\newcommand{\filetextapp}{\faFileTextO}

\newcommand{\myceil}[1]{\left \lceil #1 \right \rceil }

\newcommand{\red}[1]{\textcolor{red}{#1}}

\newcolumntype{?}{@{\hskip\tabcolsep\vrule width 1pt\hskip\tabcolsep}}

\makeatletter
\newenvironment{mcases}[1][l]
 {\let\@ifnextchar\new@ifnextchar
  \left\lbrace
  \def\arraystretch{1.2}%
  \array{@{}l@{\quad}#1@{}}}
 {\endarray\right.}
\makeatother

\newcommand{\takeaways}[1]{
\noindent \fcolorbox{gray}{white}{\begin{minipage}[c]{0.96\columnwidth}
\textbf{Takeaways}: #1
\end{minipage}}
}
\newcommand{\takeawaysNoTitle}[1]{
\noindent \fcolorbox{gray}{white}{\begin{minipage}[c]{0.96\columnwidth}
#1
\end{minipage}}
}

%% file: src/abstract.tex
\begin{abstract}
Despite the growing popularity of macOS among end users and enterprise systems, malware research has primarily focused on Windows and Android operating systems,
leaving the problem of macOS malware detection relatively unexplored.
Indeed, the specificity of the operating system and the unique characteristics of the Mach-O file format can play a fundamental role in the classification of unknown samples,
drastically increasing the detection rate.

In this work, for the first time in the literature, we employ new domain-specific features, \ie, static features specific to macOS binaries, such as embedded certificates, entitlements, persistence techniques and key system APIs,
to train a machine learning malware detector.
We perform a comprehensive experimental evaluation on a novel dataset of 41,129 samples, comprising 11,413 benign and 29,716 malicious executables, and demonstrate that our solution achieves state-of-the-art detection performance (98.50\%), outperforming all existing approaches, with an average improvement of 16\% in terms of detection rate.
We also provide an in-depth analysis of the importance of the individual features, showing that our detector effectively leverages the new domain-specific features.

Then, in order to evaluate the generalization capabilities of our detector over time, we perform a real-world evaluation on a new dataset of 9,000 fresh macOS executables.
The results show that (i) our detector maintains a very high detection rate (99.50\%), (ii) outperforms the state-of-the-art by 50\%,
and (iii) the domain-specific features are crucial for generalizing to novel malware samples, as their removal leads to a 15.92\% drop in detection performance.

Finally, we also release our dataset to the research community.
\end{abstract}

%% file: src/1-introduction.tex
\section{Introduction}
\label{sec:intro}

In recent years, the adoption of macOS among enterprises has significantly increased, and it is now part of 76\% of large US businesses. If this trend continues, macOS is expected to become the dominant operating system for enterprise endpoints by 2030~\citep{computerworld_macos_enterprise,wardle2024ArtOfMacMalware2}.
Along with its growing popularity, macOS has faced many security challenges in recent years. As reported by Kaspersky~\cite{kaspersky_blog_macos}, in 2023, a total of 17 zero-day vulnerabilities targeting macOS were discovered, including over a dozen classified as high-risk and one as critical.
The threat landscape is further exacerbated by the growing number of malware samples designed for macOS. Indeed, as reported by Moonlock Lab~\cite{moonlock_threat_report_2024}, 2024 has been characterized by a noteworthy increase in macOS malware, with a significant growth in the variety and sophistication of infostealers. 
In addition, malware authors are starting to target the \verb|arm64| architecture of new Macs, either by building \verb|arm64| binaries or by using \verb|fat| file format that supports multiple architectures~\citep{cuckoo_sentinellabs_2024}.

The increasing maturity of the malware ecosystem has prompted researchers to port existing malware detection techniques, mainly based on machine learning, to the macOS operating system.
This provides a unique opportunity to observe the evolution of these techniques and to study how the specificity of the OS affects the features and accuracy of detectors. 
In fact, while general techniques remain the same across operating systems,
the role they play in the detection of malicious samples may vary from one system to another.
For instance, the features commonly adopted to detect malicious samples in Android~\citep{arp2014drebin} differ significantly from those used for detecting Windows malware~\citep{Dambra2023DecodingTheSecretWinMalware}.

At the time of writing, existing approaches proposed in the literature still rely solely on generic static features, such as strings, byte N-grams, and the number of sections~\citep{pajouh2018intelligent,Sahoo2022,Gharghasheh2022,Chen2022MLOSXMalware,thaeler_macos_2024}.
However, the unique characteristics of the Mach-O binary format, the container for executables and libraries on macOS, and its built-in security mechanisms (such as embedded code-signing certificates and entitlements) have been largely overlooked.
Although analogous features exist on other platforms, for example, certificates in the Windows Portable Executable (PE)~\citep{Kim2017CertifiedMalware} and permissions in Android~\citep{arp2014drebin},
no prior work in macOS malware detection has systematically investigated how these macOS-specific features can be leveraged to identify malicious software and enhance the detection performance.
Additionally, it is important to understand whether the use of these features can provide additional benefits, such as enhanced generalizability or robustness to data drift and the introduction of new variants.

\vspace{0.2cm} 
\noindent \textbf{Contributions} -- In this work,
we take advantage of the emerging and rapidly evolving landscape of malware defenses for macOS to investigate the role and impact of macOS-specific features on malware detection.
In particular, we designed a number of experiments to show how unique features
of the \mbox{Mach-O} file format can enhance the performance of a machine learning-based detector.

We also observed that the absence of a large, up-to-date, publicly available dataset of benign
and malicious macOS samples has significantly limited the development of
machine learning-based solutions.
Therefore, as a key contribution of this work, we collected a new dataset containing 41,129 samples (11,413 benign and
29,716 malicious), spanning three architectures: \codeinline{x86-64}, \codeinline{arm64}, and
\codeinline{fat}.
This represents the largest dataset of macOS binaries used in the literature for malware detection.
Indeed, in comparison, the most commonly used public dataset in this area contained only 152 binaries~\citep{pajouh2018intelligent}.
To support future experiments and in the spirit of open science, we publicly release the dataset~\citep{dataset}.

The results of our experiments show that our detector, trained on macOS-specific static features, significantly outperforms existing solutions, 
with an improvement of over 16\% over the state-of-the-art solutions~\citep{pajouh2018intelligent,Sahoo2022,Gharghasheh2022,Chen2022MLOSXMalware,thaeler_macos_2024}.
Previous research works~\cite{Gibert2025WinMalware,ponte2024slifer} have also investigated deep learning methods such as MalConv~\citep{malconv2} that automatically extract significant patterns from raw byte sequences rather than relying on manual feature engineering.
However, our experiments show that MalConv is not as effective as feature-based approaches in the context of macOS malware detection.
This interesting finding highlights the importance of leveraging semantically-rich features in macOS malware detection, as they can provide more meaningful insights into the characteristics of malicious samples compared to raw byte sequences.

An in-depth analysis of the feature importance reveals two interesting results.
On the one hand, the distinctive features of macOS binaries are consistently ranked as the most important for the classification task.
On the other hand, when these features are removed, the detector is still able to achieve comparable results by relying on other, generic features.
This seems to suggest that while these specific features provide a more direct way to classify samples compared with the less semantically-rich headers information or byte N-grams,
the latter may still be sufficient to detect \textit{known} samples.

To evaluate our detector on \emph{fresh} data, \ie, temporally newer samples, we conducted a real-world assessment using 9,000 macOS binaries collected from the VirusTotal feed over a three-month period (September to November 2024).
A key finding from this second set of experiments is that macOS-specific features generalize much better to new malware samples.
In fact, while generic features were still sufficient to reliably detect known samples, performance on new data dropped by over 15\% when specific features were removed, hence demonstrating their key generalization role.

Finally, in the real-world experiment, our detector based on the full feature set continued
to achieve a high detection rate, \ie, 99.50\% at a 1\% false positive rate (FPR), even on
new samples, and continued to outperform state-of-the-art
detectors, with a remarkable 50.03\% improvement in detection rate at 1\% FPR.

%% file: src/2-background.tex
\section{Background}
\label{sec:background}

Each operating system is defined by key characteristics and unique file formats that reflect its underlying architecture and design. By leveraging this platform-specific information, we can gain critical insights into the structure and attributes of executables, enabling a more precise distinction between benign and malicious samples.
For example, researchers have leveraged the \textit{Portable Executable (PE) Rich Header} to detect Windows malware~\citep{webster2017finding}, anomalies in the \textit{ELF} file format to flag binaries designed for Linux platforms~\citep{cozzi2018understanding}, and the sequence of requested permissions as a sign of suspicious behavior for Android~\citep{arp2014drebin}.
In this paper, we focus on macOS malware, an emerging field in which researchers have so far only relied on generic features~\citep{pajouh2018intelligent,Sahoo2022,Gharghasheh2022,Chen2022MLOSXMalware,thaeler_macos_2024}, such as the number of sections and strings, but have not investigated any macOS-specific features tied to its unique file formats and architectures.

\subsection{macOS Applications and Mach-O Binaries}
\label{sec:macos_apps}

macOS applications (apps) are packaged as \emph{app bundles}~\citep{macos_app_bundle,pajouh2018intelligent}, composed of a directory tree containing various files and sub-directories. Notably, each bundle contains the main executable of the app in the \macho format, an \texttt{Info.plist} file with metadata (\eg, name, version, supported platforms by the program), a set of directories dedicated to \texttt{assets} (\eg, images, configurations), required \texttt{Frameworks} (\eg, dynamic libraries used by the app), and optional \texttt{Plugins}.

In this work, we focus on \macho executables rather than entire app bundles to extract macOS-specific indicators and build our classifiers. This choice is motivated by recent research~\citep{pajouh2018intelligent,Sahoo2022,Chen2022MLOSXMalware,Gharghasheh2022,thaeler_macos_2024}, which highlights \macho executables as the most significant artifacts for macOS malware detection. Additionally, individual binaries are the primary granularity at which analysts commonly share malicious macOS software, emphasizing the need to design features and classifiers that specifically target executables rather than app bundles. In support of this approach, we conducted a preliminary analysis of the VirusTotal (VT)~\citep{virustotal} feed over a five-month period (July--November 2023) and found that only 6\% of malware \macho executables included metadata linking them to their parent app bundles.

\vspace{0.2cm}
\takeawaysNoTitle{It is crucial to highlight that having access only to the Mach-O file of an app bundle, a very common situation for samples shared on VirusTotal, makes dynamic analysis nearly impossible and static analysis significantly harder. While malicious PE and ELF executables are typically self-contained, Mach-O binaries often depend on external files within the app's bundle, a unique characteristic that introduces a significant challenge for the analyst. The absence of these files (e.g., dynamic libraries) can prevent the sample from executing entirely or limit the analyst's ability to fully assess its capabilities, even during static analysis. For instance, if the Mach-O relies on external scripts to perform system operations, the analyst may struggle to fully understand its behavior.

}
\vspace{0.2cm}

As mentioned, the main executable of a macOS app is stored in a \macho file~\citep{wardle2022ArtOfMacMalware,aidansteelemacho_file_format}. A \macho file is a container that can represent not only executables but also object code, shared libraries, and core dumps. It consists of three main parts (see \autoref{fig:mach_o_structure}): header, load commands, and data. Every \macho file starts with a header that includes a magic number and a set of additional information that instructs the loader on how to interpret the remaining part of the file, such as the CPU architecture for which the binary has been compiled. After the header, each \macho file includes a series of load commands that describe its internal organization and instruct the loader on how to map the file contents into memory. The file is organized into segments, which represent logical regions of code or data required by the app. Each segment is further divided into sections, which contain specific types of code or data, such as executable instructions, read-only constants, or writable variables. These sections provide fine-grained organization within the broader structure of segments, allowing the operating system to handle and protect different types of data appropriately.

\input{src/macho_format}

\vspace{0.2cm}
\myparagraph{Comparison with Windows Executables.}
The \macho file format differs from the Windows PE format~\citep{win_pe} in several aspects. \macho binaries contain a single Mach header, while PE files begin with an MS-DOS stub, followed by the PE header, which includes a COFF (Common Object File Format) header and an optional header. Moreover, unlike the PE format in which data and code are organized into sections, the \macho format adopts a more complex structure consisting of segments, which are, in turn, divided into sections. However, the most distinctive feature of macOS executables is their support for \texttt{fat} binaries, \aka \emph{universal binaries}, which include code for multiple CPU architectures in the same file, thus simplifying distribution compatibility without requiring separate binaries for each architecture. A \verb|fat| executable starts with a \verb|fat| header, which includes metadata for each architecture supported by the executable. Following the header, the file contains a separate \macho binary for each architecture, each with its own header, load commands, segments, and sections. The loader uses the \verb|fat| header to select the \macho binary matching the system's CPU, ensuring a seamless execution across architectures.

\subsection{macOS Security}

macOS implements several security features to protect the system from malware and unauthorized software~\citep{apple-platform-security}. In our work, we focus on the techniques that the operating system employs to validate, load, and execute binary files.
We describe such hardening practices in the following sections and report how they enabled the extraction of a specific set of features in Section~\ref{sec:features}.

\subsubsection{Code Signing, Notarization and Gatekeeper}
They are integral components of macOS's security infrastructure, working in concert to protect users from untrusted or malicious software. \emph{Code signing}~\citep{code-signing} is the first step in this process, requiring developers to sign their applications with a Developer ID certificate issued by Apple.
This signature ensures the authenticity of every component of the bundle and ensures that its code has not been tampered with.

\emph{Notarization} adds an additional layer of security, particularly for apps distributed outside the Mac App Store. After a \macho executable is code-signed, developers can submit the entire app bundle that contains it to Apple's Notarization service, where it is scanned for security threats. It is important to note that notarization is not applicable to a single Mach-O executable but only to a complete and correctly formatted app bundle that contains it. If the app bundle passes the scan, Apple issues a ticket stapled to the app bundle's global signature, indicating that the app has been notarized.

Finally, \emph{Gatekeeper}~\citep{gatekeeper} is a security mechanism to verify that any downloaded app or plugin is signed with a valid Developer ID certificate, is notarized by Apple, and has not been tampered with since it was signed.
It permits only notarized app bundles to be executed by default. However, the user can voluntarily allow any non-notarized app to run on the system (or be deceived into doing so by the creator of the malicious software).

\subsubsection{Entitlements}
They are essentially permissions that an app needs in order to access privileged system resources or user-sensitive data, perform certain operations, or disable specific security measures~\citep{apple-entitlements}.
These permissions are encoded as key-value pairs and are embedded within the \macho code signature. Apple provides a broad range of entitlements~\citep{apple-entitlements}, which are organized into categories based on the types of resources or operations they control, such as \emph{Security}, \emph{Privacy}, and \emph{Notifications}.
Many of them fall under the \emph{Security} category, covering important components like the Hardened Runtime and App Sandbox.
However, not all entitlements are available to developers, as some are restricted to Apple's own applications.

\subsubsection{App Sandbox}
It isolates applications from critical system resources and other apps, providing an additional layer of protection against potential security breaches~\citep{apple-sandbox}. By default, the App Sandbox restricts access to sensitive system areas, such as files, hardware, and network services, ensuring that an app can only access the specific resources it needs to function. This containment minimizes the impact of security vulnerabilities within the app by preventing malicious or compromised applications from interfering with the rest of the system. Apps that require access to restricted resources must request dedicated entitlements to access the user's documents, camera, location, etc.

\subsubsection{Hardened Runtime}
It is a security feature designed to provide strong protection against exploitation by enforcing strict security policies~\citep{apple-hardened-runtime}. These policies include restrictions on dynamic code generation, loading of unsigned libraries, and preventing unauthorized debugging or code injection. By default, the Hardened Runtime is enabled, and apps that wish to perform potentially dangerous operations, which could undermine security, must request specific entitlements to bypass some of these protections. For example, an app may request entitlements for Just-In-Time (JIT) compilation or other operations that the Hardened Runtime typically blocks.

%% file: src/macho_format.tex
\begin{figure}[!t]
  \centering
  \resizebox{.75\columnwidth}{!}{
  \begin{tikzpicture}
  
  \definecolor{headercolor}{RGB}{255, 219, 187}
  \definecolor{commandcolor}{RGB}{204, 255, 204}
  \definecolor{segmentcolor}{RGB}{204, 229, 255}
  \definecolor{sectioncolor}{RGB}{255, 204, 204}
  
  \node(1)[rectangle, rounded corners=1mm, very thick, draw=black!75, fill=headercolor, minimum width=6cm, minimum height=0.6cm, anchor=west] {\textbf{Header (32 bytes)}};
  
  \node[rectangle, rounded corners=1mm, very thick, draw=black!75, fill=commandcolor, minimum width=6cm, minimum height=0.6cm, anchor=west] at (0,-0.6) {\textbf{Load Command 1}};
  \node[rectangle, rounded corners=1mm, very thick, draw=black!75, fill=commandcolor, minimum width=6cm, minimum height=0.6cm, anchor=west] at (0,-1.2) {\textbf{Load Command 2}};
  \node[rectangle, rounded corners=1mm, minimum width=6cm, minimum height=0.3cm, anchor=west] at (0,-1.8) { ... };
  \node[rectangle, rounded corners=1mm, very thick, draw=black!75, fill=commandcolor, minimum width=6cm, minimum height=0.6cm, anchor=west] at (0,-2.4) {\textbf{Load Command N}};
  
  \node[rectangle, rounded corners=1mm, very thick, draw=black!75, fill=segmentcolor, minimum width=6cm, minimum height=0.6cm, anchor=west] at (0,-3) {\textbf{Segment 1: \_\_PAGEZERO}};
  \node[rectangle, rounded corners=1mm, very thick, draw=black!75, fill=segmentcolor, minimum width=6cm, minimum height=2.3cm, anchor=west, text depth=2.3cm] at (0,-4.7) {\textbf{Segment 2: \_\_TEXT}};
  \node[rectangle, rounded corners=1mm, very thick, draw=black!75, fill=segmentcolor, minimum width=6cm, minimum height=0.6cm, anchor=west] at (0,-6.4) {\textbf{Segment 3: \_\_DATA}};
  \node[rectangle, rounded corners=1mm, minimum width=6cm, minimum height=0.5cm, anchor=west] at (0,-7.0) { ... };
  \node[rectangle, rounded corners=1mm, very thick, draw=black!75, fill=segmentcolor, minimum width=6cm, minimum height=0.6cm, anchor=west] at (0,-7.6) {\textbf{Segment M}};

  \node[rectangle, rounded corners=1mm, very thick, draw=black!75, fill=sectioncolor, minimum width=5.6cm, minimum height=0.5cm, anchor=west] at (0.2,-4.1) {\textbf{Section 1: \_\_text}};
  \node[rectangle, rounded corners=1mm, very thick, draw=black!75, fill=sectioncolor, minimum width=5.6cm, minimum height=0.5cm, anchor=west] at (0.2,-4.75) {\textbf{Section 2: \_\_cstring}};
  \node[rectangle, rounded corners=1mm, very thick, draw=black!75, fill=sectioncolor, minimum width=5.6cm, minimum height=0.5cm, anchor=west] at (0.2,-5.4) {\textbf{Section 3: \_\_const}};
  
  \end{tikzpicture}
  }
  \caption{Mach-O file format.}
  \label{fig:mach_o_structure}
\end{figure}
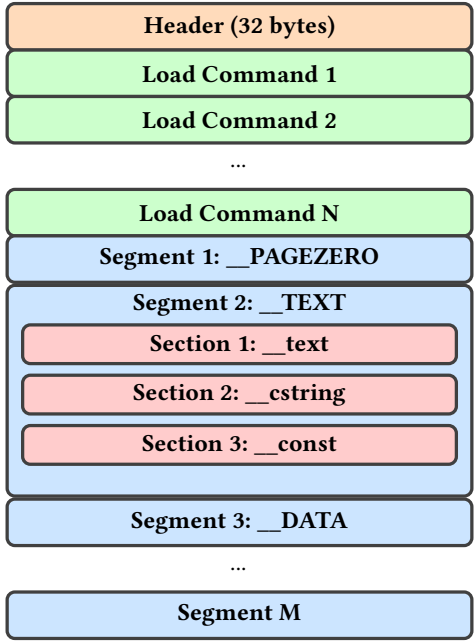

%% file: src/3-features.tex
\section{Features for macOS Malware Detection}\label{sec:features}

\begin{table}[!t]
    \footnotesize
    \centering
    \caption{Feature categories proposed in this work. N: numeric, B: boolean. Green highlighted: macOS-specific.}
    \begin{adjustbox}{max width=\columnwidth}
    \begin{tabular}{c c c c}
    \toprule
    \textbf{Category} & \textbf{Features} & \textbf{Type} & \textbf{Size} \\
    \toprule
    \multirow{12}{*}{\shortstack{Structural}}
    & Sample Total Size &  N  & 1 \\
    \cline{2-4}
    & Sample Virtual Size &  N  &  1 \\
    \cline{2-4}
    & Header Flags &  B  &  26 \\
    \cline{2-4}
    & Load Commands Histogram  &  N  &  50  \\
    \cline{2-4}
    & Min/Max/Avg Load Commands Size  & N & 3 \\
    \cline{2-4}
    & Min/Max/Avg Load Commands Entropy  & N & 3 \\
    \cline{2-4}
    & Min/Max/Avg Sections Size  & N & 3 \\
    \cline{2-4}
    & Min/Max/Avg Sections Entropy  & N & 3 \\
    \cline{2-4}
    & Min/Max/Avg Segments Size  & N & 3 \\
    \cline{2-4}
    & Min/Max/Avg Segments Virtual Size  & N & 3 \\
    \cline{2-4}
    & Min/Max/Avg Segments Entropy  & N & 3 \\
    \cline{2-4}
    & Segments Histogram  &  N  &  6 \\
    \cline{2-4}
    & Sections Histogram  &  N  &  11 \\
    \midrule
    \multirow{2}{*}{\shortstack{Byte}}
    & Byte Histogram &  N  &  256  \\
    \cline{2-4}
    & Bytes N-grams &  N & 128 \\
    \midrule
    String & \makecell{Count of strings representing common IoC}
    &  N & 5 \\
    \midrule
    \multirow{2}{*}{\shortstack{Packing}}
    & \makecell{Presence of sections with suspicious names} &  B  &  1  \\
    \cline{2-4}
    & Presence of segments ($>$ 20\%) with high-entropy data &  B & 1 \\
    \midrule
    \rowcolor{green!40}
    API & Presence of common macOS API &  B  &  4009  \\
    \midrule
    \rowcolor{green!40}
    Entitlements & Presence of common and security-related entitlements &  B  &  55  \\
    \midrule
    \rowcolor{green!40}
    Persistence  & Check for login and launch items &  B  &  2  \\
    \midrule
    \rowcolor{green!40}
    Certificates  &  Status of certificate chain &  N  &  5 \\
    \bottomrule
    \end{tabular}
    \end{adjustbox}
    \label{table:features}
\end{table}

In our detector, each Mach-O executable is represented by a 4,578-dimensional feature vector.
As summarized in \autoref{table:features}, we group each feature into eight main categories, defined according to the source used to extract them: file structure, bytes, strings, packing, APIs, entitlements, persistence techniques, and certificates.
For each feature, the table shows the data type, \ie, numeric (float) or boolean, as well as its size, \ie, the number of values used to represent the feature.
These features have been designed to cover the set of features commonly adopted for Windows malware detection~\citep{Anderson2018Ember,Dambra2023DecodingTheSecretWinMalware}, and are complemented with platform-dependent features dedicated to capture the peculiarities of the \macho file format, such as entitlements, certificates, and those related to persistence.
It is worth noting that, even though the features that are exclusively specific to macOS binaries represent only 62 out of 4,578 features, in the experiments we will show that they are valuable for the classification of macOS samples (see Subsection~\ref{subsec:feat_importance}).
The remainder of this section describes in more detail how we adapted existing indicators to macOS binaries and how we computed macOS-specific features.

\vspace{0.3cm}
\noindent
\myparagraph{Structural features.}
This set includes commonly used features associated with structural properties of a binary file, which have been extended from the literature on Windows malware detection~\citep{Dambra2023DecodingTheSecretWinMalware,Anderson2018Ember}. Specifically, in addition to common features such as total and virtual size of the sample, we consider 50 of the most common load commands that can be present in a Mach-O file~\citep{aidansteelemacho_file_format}, and count how many times each of them is included in the binary (\emph{Load Commands Histogram} in \autoref{table:features}).
Furthermore, we compute additional statistical features, such as max, min, and average size and entropy of sections, load commands, and segments present in the binary.
Moreover, we count the number of common segments, namely \codeinline{__TEXT}, \codeinline{__DATA}, \codeinline{__PAGEZERO}, \codeinline{__OBJC}, \codeinline{__IMPORT}, \codeinline{__LINKEDIT}, and sections, \ie, \codeinline{__text}, \codeinline{__data}, \codeinline{__bss}, \codeinline{__dylb}, \codeinline{__cstring}, \codeinline{__const}, \codeinline{__la_symbol_ptr}, \codeinline{__literal4}, \codeinline{__literal8}, \codeinline{__jump_table}, and \codeinline{__pointers}.
These features are summarized by the \emph{Segments Histogram} and \emph{Sections Histogram} in \autoref{table:features}.
Finally, to support \verb|fat| binaries, we compute the average of the numeric features among the executables contained in the \verb|fat| binary, while for boolean features (such as \emph{Header Flags}) we apply the logical \emph{OR} operation, \ie, if at least one of the binaries has the feature, the resulting feature is set.

\vspace{0.3cm}
\noindent
\myparagraph{Byte-based features.}
These features represent statistical properties of the byte sequence that composes the binary. We compute them by respectively counting the occurrence of each byte (i.e., byte histogram) and by extracting byte N-grams (N=2 in our experiments) with a hashing technique applied to map these patterns into a fixed-dimensional space of 128 features for efficient representation, following the approach in EMBER~\citep{Anderson2018Ember}.

\vspace{0.3cm}
\noindent
\myparagraph{String-based features.}
This category captures the presence of common strings associated with \gls*{IOC}. Specifically, we extract five classes of strings that are more likely to be relevant for malware detection, namely network resources (IPs and URLs), filesystem paths, base64-encoded strings, imported libraries, and functions' names.
As for the strings related to network resources, filesystem paths, and base64-encoded strings, we leverage regular expressions to identify them.
On the other hand, libraries and functions' names have been extracted using the LIEF library~\citep{lief}.
To reduce the noise introduced by the high number and variety of strings in the binaries, we compute the number of strings belonging to each class and obtain five numerical features.

\vspace{0.3cm}
\noindent
\myparagraph{Packing-based features.}
Packing is a widely used technique by malware authors to obfuscate malicious code and hinder static analysis efforts, and previous research has shown that macOS malware authors also leverage it for the same purpose~\citep{wardle2024ArtOfMacMalware2}. Indeed, some well-known samples, such as \emph{oRat}, \emph{IPStorm}, \emph{ZuRu}, \emph{Coldroot} and \emph{OceanLotus}, use common packers such as \texttt{UPX}~\citep{wardle2024ArtOfMacMalware2}. Our work is the first to incorporate packing-based features into a macOS malware detector, in particular by using two features that reflect the presence of packing. The first checks for the presence of sections with names related to the \verb|UPX| packer, such as \codeinline{__XHDR}, \codeinline{UPX_DATA}, and \codeinline{upxTEXT}. We focused on \verb|UPX| because it is one of the most common packers for \macho binaries used by macOS malware authors. Indeed, researchers from MoonLock Labs~\citep{moonlock_threat_report_2024} found that $~$26\% of the packed malware samples analyzed in 2024 were packed with \verb|UPX|, making it the primary choice for Mach-O compression.
The second feature checks if the binary includes more than 20\% of segments with high-entropy data (\ie, if the entropy is higher than 7), which indicates that the binary is likely packed or compressed. This feature is inspired by the approach of \texttt{pefile}~\citep{pefile} to detect packing in Windows malware samples and is also adopted in~\cite{wardle2024ArtOfMacMalware2} for macOS binaries.

\newpage
\myparagraph{API-based features.}
This category of features aims to capture how the samples use common macOS APIs.
Notably, while macOS APIs have previously been identified as good indicators to guide manual analysis of macOS malware~\citep{wardle2022ArtOfMacMalware, manna2021}, none of the existing works have used them as features for machine learning-based detection.
In this work, we show that API calls are effective for macOS malware detection when used as features (see \autoref{subsec:feat_importance}), thereby supporting analogous findings in the Windows malware detection literature~\citep{dl_malware_api}, where APIs have proven effective in capturing the malicious behavior.
Specifically, we first leverage the official Apple documentation~\citep{apple-docs-generic} to identify the most common frameworks used in macOS applications, such as \codeinline{AppKit}, \codeinline{CoreFoundation}, \codeinline{SystemConfiguration}, \codeinline{Kernel}, as well as those related to security and privacy, such as \codeinline{Security}.
To this end, we identified a total of 35 frameworks.
Next, we remove all the APIs that are included in fewer than 10 samples (\ie, 0.02\% of the samples) in the dataset since they are not very representative of the common behavior of the samples and generally represent outliers obtained when extracting the imported APIs using the LIEF library~\citep{lief}.
For each framework, we identify the 400 most common APIs\footnote{If a framework has fewer than 400 APIs, we include all available ones.}, setting this threshold based on our finding that the average number of APIs per framework in our dataset is around 400.
This process results in a total of 4,009 unique APIs, each represented by a boolean feature.

\vspace{0.3cm}
\noindent
\myparagraph{Entitlements-based features.}
This category, specific to macOS, captures the presence of common entitlements.
Specifically, we create an initial list of entitlements by leveraging the official Apple documentation~\citep{apple-entitlements} to identify all the main security-relevant entitlements, such as those related to App Sandbox and Hardened Runtime.
Then, to complement the above list with common entitlements used by the samples in the dataset, we select those present in at least 1\% of the samples (\ie, ~411) and add those not already in the initial list.
Finally, we compute a boolean feature for each of the resulting 55 entitlements that we obtained through our selection process.

\vspace{0.3cm}
\noindent
\myparagraph{Persistence-based features.}
We leverage two boolean features to capture two common persistence techniques used in macOS: login items and launch items.

Login items are applications that start automatically at user login and run within the user's desktop session by inheriting the user's permissions.
Based on previous research~\citep{wardle2022ArtOfMacMalware,wardle2024ArtOfMacMalware2}, we check for the usage of common APIs such as \codeinline{LSSharedFileListCreate}, \codeinline{LSSharedFileListInsertItemURL}, \codeinline{SMLoginItemSetEnabled}, and \codeinline{registerAndReturnError}, specifically designed to manage and interact with applications that are set to launch automatically during user login.
This persistence mechanism has been observed in several macOS malware families, such as \emph{Kitm}, \emph{NetWire}, and \emph{WindTail}~\citep{wardle2022ArtOfMacMalware}.

Launch items are, instead, persistence mechanisms designed for service executables, such as software updaters, background processes, and daemons.
They can be classified into launch agents and launch daemons.
While the former run once after login with standard user permissions and may interact with the user session, the latter are non-interactive daemons launched before user login and run with \verb|root| permissions.
As explained in~\citep{wardle2022ArtOfMacMalware,wardle2024ArtOfMacMalware2}, a common way to achieve persistence by malware samples is to create at runtime a property list (\emph{.plist}) file in \codeinline{/Library/LaunchAgents} or \codeinline{/LaunchAgents} for launch agents, or \codeinline{/Library/LaunchDaemons} for launch daemons, with the \codeinline{<key>RunAtLoad</key>} set to \verb|true|, which tells the macOS system to start the launch item automatically.
This technique has been observed in several macOS malware families, such as \emph{AppleJeus}, \emph{DazzleSpy}, and \emph{EvilQuest}~\citep{wardle2022ArtOfMacMalware}.
To detect this persistence method in our samples, we check for the presence of the \codeinline{<key>RunAtLoad</key>} string within the executable.
However, it is worth noting that some malware, such as \emph{EvilQuest}, can use obfuscation techniques to hide the presence of the embedded \emph{.plist} file~\citep{wardle2022ArtOfMacMalware}, bypassing our simple feature extraction technique.

\vspace{0.3cm}
\noindent
\myparagraph{Certificates-based features.}
This set of features aims to summarize the status of the certificate chain included in the binary (if any).
To achieve this, we map five boolean features to the potential states in which a certificate chain embedded in a binary may be found (see \autoref{table:features}):

\begin{itemize}[leftmargin=*]
    \item \textit{Certificate chain found:} true if the binary includes a certificate chain (\ie, if the certificates are included in the \emph{Code Signature} load command).
    \item \textit{Certificate chain expired:} true if at least one of the certificates in the chain is expired.
          In this case, we consider a certificate expired if the date of the analysis (\ie, 2nd December 2024) is after the expiration date of the certificate.
          Despite being time-dependent, this feature is included both for completeness and because it is generally required for certificate validation.
    \item \textit{Certificate chain self-signed:} true if at least one of the certificates in the chain is self-signed.
    \item \textit{Certificate chain revoked:} true if at least one of the certificates in the chain is revoked.
    \item \textit{Certificate chain validated:} true if the root certificate is signed by the Apple Root Certification Authority (CA).
\end{itemize}

%% file: src/4-dataset.tex
\section{Dataset}\label{sec:dataset}
As discussed in Section~\ref{sec:intro}, one of the main obstacles that has severely limited 
macOS malware research is the lack of a large, up-to-date, and publicly available dataset of macOS binaries. 
What researchers have previously created for their experiments is too small to effectively train machine learning models. 
For instance, the dataset created and published by Pajouh \etal~\cite{pajouh2018intelligent} in 2018 and used in subsequent studies~\citep{Sahoo2022,Chen2022MLOSXMalware,Gharghasheh2022} until 2022 contains only 152 binaries. 
Other datasets, such as those curated by Walkup \etal~\cite{Walkup2014MacMalwareStatic} in 2014 and Thaeler \etal~\cite{thaeler_macos_2024} in 2024, have never been publicly released, further limiting access to comprehensive resources for macOS malware analysis.

To tackle this challenge and support the training and evaluation of our machine learning models, we assembled a new dataset of macOS executables by combining samples obtained from a variety of sources. 
Our new dataset includes 41,139 Mach-O executables, with 11,413 goodware samples and 29,716 malware samples, and all samples were carefully processed to ensure its suitability for large-scale analysis and machine learning applications. 
To further support the research community, we make our dataset publicly available~\citep{dataset}.

\subsection{Sample Sources}

Goodware samples were collected from the following three sources.
\begin{itemize}[leftmargin=*]
\item{\textbf{macOS installation.} We extracted 1,239 Mach-O executables from a macOS Sonoma 14.5 installation on an Apple M1 MacBook Pro. Specifically, we collected preinstalled system binaries and applications from the default installation paths, including \verb|/bin|, \verb|/sbin|, \verb|/usr/bin|, \verb|/usr/sbin|, \verb|/Applications|, and $\sim$\verb|/Applications|.}

\item{\textbf{Homebrew.} We extracted 9,843 executables from macOS applications available through the Homebrew package manager~\citep{homebrew}, filtering out those whose status is either \emph{deprecated}, \emph{disabled}, or \emph{outdated}.}

\item{\textbf{Open-source macOS apps.} We collected the remaining 1,045 executables from several open-source macOS applications available on GitHub~\citep{open-source-mac-os-apps}.} 
\end{itemize}

\noindent Malware samples were collected from a variety of sources.
\begin{itemize}[leftmargin=*]
\item{\textbf{Objective-See dataset.} A curated dataset of macOS malware samples~\citep{objectivesee} maintained by Patrick Wardle and
    also used in previous research studies~\citep{pajouh2018intelligent,thaeler_macos_2024}, which included 180 samples at the time we accessed it.}
\item{\textbf{MalwareBazaar.} We downloaded all the samples from MalwareBazaar~\citep{malwarebazaar} tagged with \verb|macho| and detected as malicious by at least 5 antivirus (AV) engines.
    At the time of the download, we found 90 samples.}
\item{\textbf{Virus Samples Team dataset.} This dataset includes 103 samples that
	we downloaded from their GitHub repository~\citep{macvirusdataset}.}
\item{\textbf{VirusShare.} We used 2,910 malware samples from VirusShare~\citep{virusshare}.
    Since the platform does not provide any API to filter for macOS samples, we leveraged the reports and information provided in the MalDICT~\citep{joyce2023maldict} paper by selecting all samples available on VirusShare tagged as \verb|macos|.
    Moreover, for each selected sample, we collected the related report to filter out all samples detected by less than 5 AV engines.}
\item{\textbf{VirusTotal.} 
    We monitored the VT feed from July 1st to November 13th 2023 and retained samples with the
    \verb|mac| or \verb|macho| tag, which were detected by at least 5 AV engines.
    This resulted in 28,380 additional macOS malware samples.}
\end{itemize}

\subsection{Sample Extraction}
After collecting goodware and malware samples from different sources, we processed them to create a homogeneous dataset. 
Since the samples can be distributed using different archive file formats, such as \verb|.dmg| and \verb|.zip|, we implemented a fully automated pipeline that extracts the contents of the archive based on the file format and checks if it contains any Mach-O executable. After the extraction, we filtered out all binaries that are not Mach-O executables (e.g., dynamic libraries) and those compiled for a CPU architecture other than \verb|x86-64| or \verb|arm64| (e.g., \verb|PowerPC|). We also excluded all the \macho samples intended for iOS and WatchOS by checking the \emph{platform} field in the \emph{Build Version} load command (\verb|LC_BUILD_VERSION|). Finally, we processed each sample using a LIEF-based script~\citep{lief} and removed those that could not be properly parsed and those lacking load commands or sections.

As for the goodware samples, we implemented an extra validation step by using VirusTotal to verify that none were detected as malicious by any of the AV engines available on the platform.

As a result of these processing steps, our dataset consists of 41,129 samples, including 11,413 goodware and 29,716 malware samples.

\begin{table}[!t]
    \centering
    \footnotesize
    \caption{Samples distribution by CPU architecture.
    Percentages in the \emph{Total} row refer to all samples, while the others are computed relative to each label.}
    \begin{tabular}{c c c c c}
          \toprule
          Label  &  \verb|x86-64|  &  \verb|arm64|  &  \verb|fat|  &  Total  \\
          \toprule
          Malware    &  27,339 (92\%)  &  1,812 (6\%)  &  565 (2\%)      &  29,716  \\
          Goodware   &  3,813 (33\%)   &  2,104 (18\%)  &  5,496 (48\%)  &  11,413  \\
          \midrule
          Total  &  31,152 (76\%)  &  3,916 (9\%)   &  6,061 (15\%)  &  41,129  \\
          \bottomrule
    \end{tabular}

    \label{tab:arch_distribution}
\end{table}

\begin{table}[!t]
  \centering
  \caption{Families containing at list 1\% of the malware samples.}
  \begin{adjustbox}{max width=\columnwidth}
  \begin{tabular}{c c c c c c c c}
        \toprule
        \verb|bundlore|  &  \verb|adload|  &  \verb|pirrit|  &  \verb|jailbreak| & \verb|evilquest| &  \verb|lador|  &  \verb|genieo|  &  \verb|stealer| \\
        \midrule
        \makecell{9,484 \\ (33.84\%)}  &  \makecell{7,131 \\ (25.44\%)}  &  \makecell{2,934 \\ (10.47\%)}  &  \makecell{582 \\ (2.08\%)} & \makecell{464 \\ (1.66\%)}  &  \makecell{454 \\ (1.62\%)}  &  \makecell{438 \\ (1.56\%)}  &  \makecell{415 \\ (1.48\%)} \\
        \bottomrule
  \end{tabular}
  \end{adjustbox}
  \label{tab:family_distribution}
\end{table}

\subsection{Dataset Analysis}
\autoref{tab:arch_distribution} shows the distribution of the samples by CPU architecture: 76\% of the samples are compiled for \verb|x86-64|, 9\% for \verb|arm64| CPUs, and 15\% are \verb|fat| binaries, \ie, they support both the \verb|x86-64| and \verb|arm64| architectures. 
It is noteworthy how goodware and malware samples are distributed differently across the target architectures. 
In fact, the majority of the benign samples (5,496, corresponding to 48\% of the goodware dataset) are \verb|fat| binaries, while the distribution of malware samples is heavily skewed towards the \verb|x86-64| architecture, with 27,344 samples (92\% of all malicious files). 
This result confirms that malware authors continue to target the \verb|x86-64| architecture~\citep{sentinelone_adload_2024,bluenoroff_sentinellabs_2024}, suggesting that \verb|x86-64| remains prevalent in the desktop Apple ecosystem. 

\autoref{tab:family_distribution} shows the distribution of malicious samples by family according to the most common label provided by VT. We considered only the families containing at least 1\% of the total number of malware samples in the dataset. This results in eight families, which cover about 80\% of all malware samples. 
The family distribution of our dataset, which mainly reflects the prevalence of real-world families collected in the wild, is highly imbalanced. Indeed, the top three families, namely \texttt{bundlore}~\citep{bundlore_mitre} (33.84\%), \texttt{adload}~\citep{sentinelone_adload_2021,sentinelone_adload_2024} (25.44\%), and \texttt{pirrit}~\citep{pirrit_paloalto_2023} (10.47\%), account for 70\% of the malware samples.
Due to the low number of significant families and the high imbalance of the remaining ones, in this work we focused on binary classification. Nevertheless, we also performed experiments with family classification and included the results in Appendix~\ref{sec:family_classification}.

\vspace{0.2cm}
\subsubsection{Presence of packing}
Our dataset also allowed us to perform a large-scale analysis of packing in macOS malware, which has never been studied before.
To identify packed samples, we used two static features commonly associated with packing~\cite{wardle2024ArtOfMacMalware2}:
the presence of suspicious section names related to UPX and high-entropy data sections. As shown in \autoref{tab:analysis_packing}, we found 4,362 ($\sim$2\% of the total) potentially packed samples, of which
4,069 are malicious ($\sim$14\% of the malware set) and 293 ($\sim$2\% of the goodware set) are benign. Among the potentially packed malware samples, the
majority (3,938) are \verb|x86-64| binaries, followed by 122 \verb|arm64| binaries and 9 \verb|fat| binaries.
As for the potentially packed goodware samples, we found that most of them (139) are \verb|x86-64| binaries, while 50 are \verb|arm64| and 104 are \verb|fat| binaries.

\begin{table}[!t]
    \footnotesize
    \centering
    \caption{Samples distribution for packing-based features.}
    \begin{adjustbox}{max width=\columnwidth}
    \begin{tabular}{c c c c c c c}
        \toprule
        \textbf{Label} & \multicolumn{3}{c}{\textbf{Suspicious Section Names}} & \multicolumn{3}{c}{\textbf{High Entropy Sections}} \\
        \cmidrule(lr){2-4} \cmidrule(lr){5-7}
        & \verb|x86-64| & \verb|arm64| & \verb|fat| & \verb|x86-64| & \verb|arm64| & \verb|fat| \\
        \toprule
        Malware  & 21  &  0  &  0  &  3917  &  122  &  9   \\
        Goodware  & 1   &  0  &  0  &  138   &  50   & 104  \\
        \midrule
        Total & 22 (0.05\%) & 0 (0\%) & 0 (0\%) & 4055 (9.85\%) & 172 (0.42\%) & 113 (0.27\%) \\
        \bottomrule
    \end{tabular}
    \end{adjustbox}
    \label{tab:analysis_packing}
\end{table}

\vspace{0.2cm}
\takeawaysNoTitle{
Packing is not very prevalent among macOS malware executables, with around 10\% of the samples in our dataset showing clear signs of packing. As is the case for other operating systems, we also identified potentially packed benign software. Moreover, it is interesting to observe that the proportion of potentially packed samples is considerably higher for \texttt{x86-64} binaries, likely because common packing techniques are more mature for this architecture.
}
\vspace{0.2cm}

\begin{figure}[!t]
    \centering
    \includegraphics[width=\columnwidth]{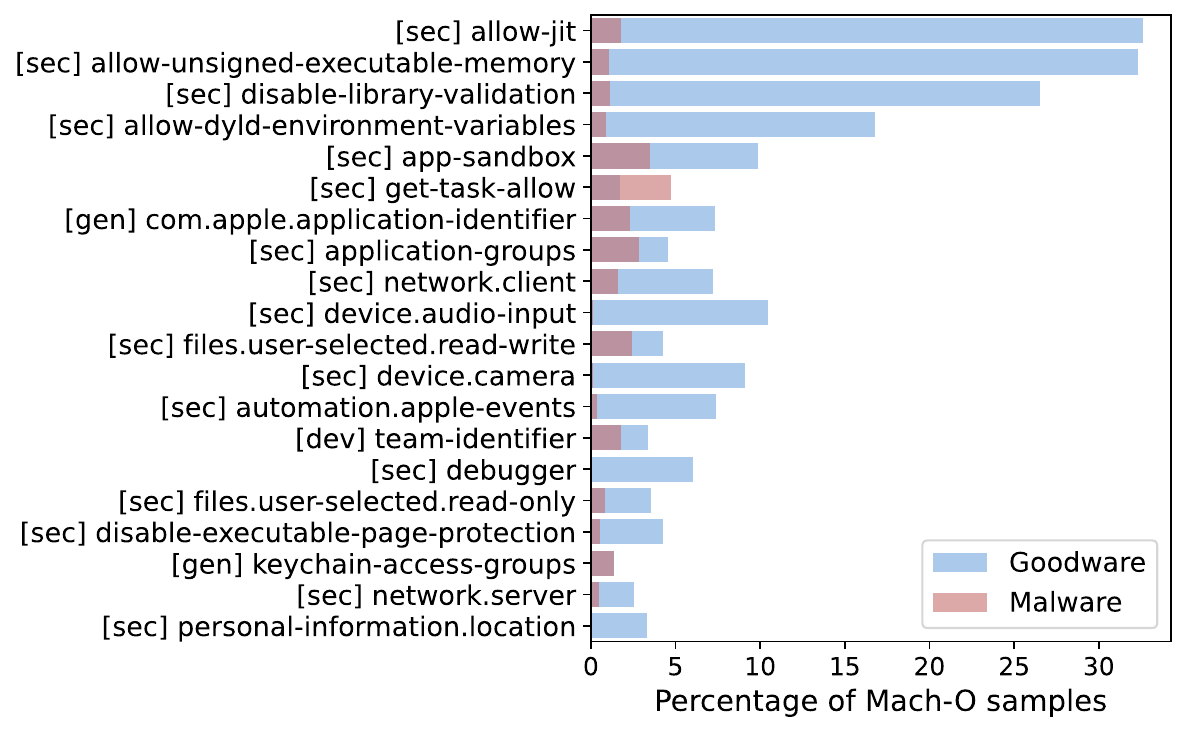}
    \caption{
    Top-20 entitlements.
    \codeinline{[sec]}, \codeinline{[dev]} and \codeinline{[gen]} prefixes represent the entitlements' category: security-related, developer-related and generic one, respectively.
    }
    \label{fig:analysis_entitlements}
\end{figure}

\begin{figure*}[!t]
    \centering
    \includegraphics[width=\textwidth]{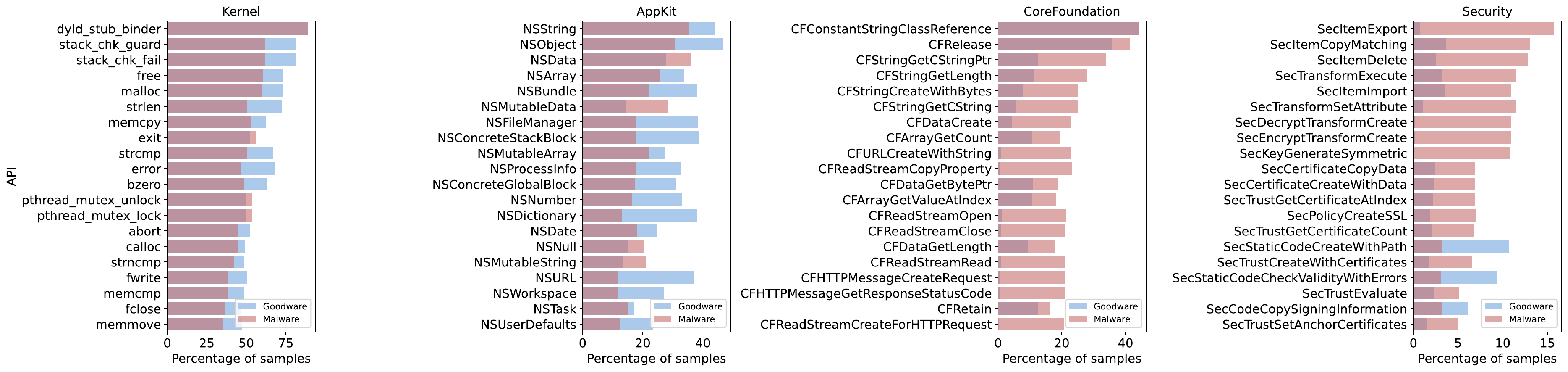}
    \caption{Analysis of the most common macOS APIs.}
    \label{fig:analysis_api}
\end{figure*}

\subsubsection{Differences in macOS APIs usage}
To further analyze the API usage by the samples in the dataset, we considered the top-4 frameworks containing the highest number of APIs: \codeinline{Kernel} (including the C standard and BSD libraries), \codeinline{AppKit} \& \codeinline{Foundation} (referred to as \codeinline{AppKit} hereafter), \codeinline{CoreFoundation} (including \codeinline{CFNetwork}), and \codeinline{Security}. \autoref{fig:analysis_api} shows the distribution of the top-20 imported APIs per framework, ordered by their total usage across both goodware and malware.
The bars report the percentage of goodware and malware samples that import each API, computed over the total number of goodware and malware samples, respectively.
For the \codeinline{Kernel} and \codeinline{AppKit} frameworks, we observed that the top APIs are more frequently used by goodware than malware. This is expected, given their role in standard system functionality and user interface development.
In contrast, the APIs belonging to the \codeinline{CoreFoundation} and \codeinline{Security} frameworks, which are respectively adopted for low-level system and cryptographic operations, are used more frequently by malware.

\vspace{0.2cm}
\takeawaysNoTitle{
Malware samples tend to use a higher number of low-level APIs from \codeinline{CoreFoundation}, while goodware ones tend to rely on high-level APIs. In addition, malware samples employ more security-related APIs, especially those that access the keychain (e.g., \codeinline{SecItemExport}) or perform cryptographic operations (e.g., \codeinline{SecEncryptTransformCreate}).
}
\vspace{0.2cm}

\subsubsection{Entitlements distribution}
\label{subsec:ent}
In~\autoref{fig:analysis_entitlements}, we show the distribution of the top-20 entitlements and the category of the framework containing them to assess potential differences between benign and malicious samples.
Seventeen of the top-20 entitlements are security-related and are far more commonly used by goodware rather than malware.
For instance, the most common entitlement, \codeinline{allow-jit}, which is related to the Hardened Runtime and allows the binary to execute JIT-compiled code, is used by 32.6\% of goodware but only 1.8\% of malware.
Moreover, among \verb|arm64| samples, 99.13\% of goodware samples use this entitlement, compared to just 0.87\% of malware samples.
This further confirms that this entitlement seems to be a distinctive feature of arm64 goodware samples.
Similarly, the second most common entitlement, \codeinline{allow-unsigned-executable-memory}, is used by 32.3\% of goodware and only 1.1\% of malware.

\vspace{0.2cm}
\takeawaysNoTitle{
The presence of security-related entitlements such as \codeinline{allow-jit} is a very distinctive feature of goodware samples.
As we will unveil in our results, the presence of such entitlements is an important feature for distinguishing between goodware and malware.
}
\vspace{0.2cm}

\begin{table}[!t]
    \footnotesize
    \centering
    \caption{Samples distributions for persistence mechanisms.}
    
    \begin{tabular}{c c c c c c c}
        \toprule
        \textbf{Label} & \multicolumn{3}{c}{\textbf{Login Items}} & \multicolumn{3}{c}{\textbf{Launch Items}} \\
        \cmidrule(lr){2-4} \cmidrule(lr){5-7}
        & \verb|x86-64| & \verb|arm64| & \verb|fat| & \verb|x86-64| & \verb|arm64| & \verb|fat| \\
        \toprule
        Malware  & 262   &  20  &   29  &  432  &  120   &  36  \\
        Goodware  & 103   &  10  &  269  &    8  &    8   &  25  \\
        \midrule
        Total & 365 & 30 & 298 & 440 & 128 & 61 \\
        \bottomrule
    \end{tabular}

    \label{tab:analysis_persistence}
\end{table}

\subsubsection{Differences in persistence mechanisms}
We found 1,322 samples ($\sim$3\% of the dataset) that leverage persistence techniques based on the persistence-related features described in Section~\ref{sec:features}: 899 are malicious (3\% of malware) and 423 are benign (3.7\% of goodware). \autoref{tab:analysis_persistence} presents their distribution by persistence mechanism (login items vs. launch items) and architecture.
As for login items, no major difference is observed between goodware and malware overall.
However, looking at the single architecture distributions, we can see that this persistence technique is tightly correlated to the CPU architecture: it is used more often by malware samples based on the \verb|x86-64| architecture (262 samples), while it is more commonly used by goodware samples that are \verb|fat| binaries (269 samples).
On the other hand, launch items have a different trend.
Indeed, this persistence mechanism is used more often by malware (588) than by goodware (41) across all architectures.
Specifically, it is widely used by malware samples based on the \verb|x86-64| architecture (432 samples) and less frequently by the other architectures (120 and 36 samples for \verb|arm64| and \verb|fat| binaries, respectively).

\vspace{0.2cm}
\takeawaysNoTitle{
Persistence is not widely used in the dataset, with only 3\% of samples employing such techniques. The presence of login items depends on both the sample's nature and its architecture, while the presence of launch items is more indicative of malicious behavior, being more common in malware samples regardless of the architecture.
}
\vspace{0.2cm}

\subsubsection{Analysis of the certificate status}
\label{subsec:certs}
According to their certificate chain status, we divided the samples into the following nine groups:
\begin{itemize}[leftmargin=*]
    \item \texttt{nocert}: no certificate chain found.
    \item \texttt{novalid}: certificate chain found but not validated by Apple.
    \item \texttt{revoked valid}: certificate chain found and validated, but at least one certificate is revoked.
    \item \texttt{self-signed}: certificate chain found, with at least one certificate being self-signed.
    \item \texttt{self-signed expired}: certificate chain found, with at least one certificate being both self-signed and expired.
    \item \texttt{expired novalid}: certificate chain found, not validated, and with at least one certificate expired.
    \item \texttt{expired valid}: certificate chain found and validated, with at least one certificate expired.
    \item \texttt{expired revoked valid}: certificate chain found and validated, with at least one certificate expired and at least one that is revoked.
    \item \texttt{valid}: certificate chain found and validated by Apple.
\end{itemize}

The results, reported in \autoref{tab:certificate_status}, show that the majority ($\sim$83\%) of malware samples do not include a certificate chain, while most goodware samples ($\sim$65\%) include a certificate chain validated by Apple. Interestingly, roughly 3\% of malware samples include a certificate chain validated by Apple, which is neither expired nor revoked. The presence of self-signed certificates is very low in both goodware (0.26\%) and malware (0.19\%), even when considering samples with certificate chains that are both expired and self-signed (0.13\% for goodware and 0.04\% for malware).

\begin{table*}[!t]
    \centering
    \footnotesize
    \caption{Percentages of sample with different certificate status.}
   
    \begin{tabular}{c c c c c c c c c c}
        \toprule
        \textbf{Label} & \codeinline{nocert} & \makecell{\codeinline{revoked} \\ \codeinline{valid}} &  \codeinline{self-signed} & \makecell{\codeinline{self-signed} \\ \codeinline{expired}} & \makecell{\codeinline{expired} \\ \codeinline{novalid}} & \makecell{\codeinline{expired} \\ \codeinline{valid}} & \makecell{\codeinline{expired} \\ \codeinline{revoked} \\ \codeinline{valid}} & \codeinline{novalid} & \codeinline{valid} \\
        \toprule
        Malware    & 83.44   &  4.32  &  0.19  &  0.04  &  0.01   &  5.66   &  3.43  &  0.03  &  2.87   \\
        Goodware   & 14.65   &  0.12  &  0.26  &  0.13  &  0.00   &  19.59  &  0.07  &  0.00  &  65.17  \\
        \bottomrule
    \end{tabular}

    \label{tab:certificate_status}
\end{table*}

\vspace{0.2cm}
\takeawaysNoTitle{
As we will reveal in our results, the status of the certificate chain, in particular the presence of certificates and whether it is validated by Apple, is a key feature for distinguishing between goodware and malware samples.
In addition, security analysts can easily interpret this information, making it highly valuable for explainability and for identifying the root cause of a detection.
}
\vspace{0.2cm}

%% file: src/5-experiments.tex
\section{Experiments \& Results}
\label{sec:exp_results}

In this section, we introduce the experimental setup, describe the experiments conducted to train and test our detector based on the proposed features, and compare it with other state-of-the-art approaches.
This is followed by a discussion of the results. Specifically, the evaluation aims to address the following research questions:

\vspace{0.1cm}
\myparagraph{(RQ.1) Feature Effectiveness} -- 
Do the proposed macOS-specific features enhance detection performance when compared to existing state-of-the-art approaches?

\myparagraph{(RQ.2) Feature Importance} -- 
Are the proposed macOS-specific features effectively leveraged by our detector?

\vspace{0.1cm}
\myparagraph{(RQ.3) Feature Robustness} -- 
Do the macOS-specific features generalize to new variants?
Which role do they play in the detection of new samples collected in the wild?

\vspace{0.1cm}
\myparagraph{(RQ.4) Real-World Accuracy} -- 
How does a detector equipped with our features compare with state-of-the-art macOS models 
in a real-world assessment?

\subsection{Experimental Setup}
\label{sec:exp_setup}

All experiments were conducted on an Ubuntu 22.04.6 LTS server equipped with an Intel Xeon Platinum 8160 CPU @ 2.10 GHz (64 cores) and 256 GB of RAM.

We performed a comprehensive benchmarking by comparing our set of features with those proposed in the literature by~\cite{thaeler_macos_2024} and~\cite{pajouh2018intelligent}, the latter of which has been widely adopted in subsequent works~\citep{Sahoo2022,Chen2022MLOSXMalware,Gharghasheh2022}. Hence, we refer to all these as \emph{Pajouh-based} in the following.
We evaluated several tree-based machine learning models, namely Decision Tree (DT), Random Forest (RF)~\citep{breiman2001rf}, and XGBoost~\citep{xgboost}, since they are widely adopted in the literature dedicated to Windows malware detection~\citep{Dambra2023DecodingTheSecretWinMalware,Anderson2018Ember,Gibert2025WinMalware}, have been proven to outperform deep learning models on tabular data, as in our case,~\citep{tree_based_models_neurips}, and offer many advantages such as robustness to outliers and explainability~\citep{molnar2022}.
All the tree-based models were implemented and trained by using the following Python packages: \verb|scikit-learn| v1.5.0~\citep{sklearn}, \verb|xgboost| v2.1.0, and \verb|crepes| v0.7.1~\citep{crepes} for model calibration.
To tune the models' hyper-parameters (see \autoref{tab:hyperparams} in Appendix~\ref{appendix:hyperparameters}) in line with the state-of-the-art~\citep{Trizna2024Nebula}, we performed a grid search based on 5-fold cross-validation (CV) on the training set to ensure a fair evaluation~\citep{raschka2018}. Since this resulted in five different models, all the reported results are the mean values across the five models, each one evaluated on its corresponding test set obtained from the CV split.
We additionally tested the models' generalization capabilities on a temporally newer sample distribution consisting of a collection of real-world macOS benign and malicious samples collected over a 3-month period.
In addition, we evaluated MalConv~\citep{malconv2}, an end-to-end deep learning model for raw byte sequences classification, widely adopted for Windows malware detection~\citep{kozak2024updatingwindowsmalwaredetectors,ponte2024slifer,Gibert2025WinMalware}. It was trained using the default architecture and hyper-parameters described in~\citep{malconv2}.
We evaluated the performance of all the models for all the target CPU architectures, as well as in the architecture-agnostic scenario.
To this end, based on our analysis in \autoref{sec:dataset}, we split our dataset by CPU architecture and created four different \emph{benchmarking datasets} (see \autoref{tab:arch_distribution}): \verb|x86-64|, \verb|arm64|, \verb|fat|, and \verb|multi-arch|, the latter of which includes all the binaries regardless of the CPU architecture.
For completeness, the family classification performance of our solution is reported in Appendix~\ref{sec:family_classification}.

\begin{table}[!t]
    \centering
    \footnotesize
    \caption{TPR @ 1\% FPR of the evaluated models. In bold the best results for each feature set, used to compute the average improvement
    as the average of the relative improvements of our best model (\texttt{XGBOOST}) compared to the best-performing model for the other feature sets as
    $(\text{our} - \text{other}) \, / \, \text{other} \times 100$.}
    \begin{tabular}{c c c c c c}
        \toprule
        Features & Model &  \verb|x86-64|  &  \verb|arm64|  &  \verb|fat|  &  \verb|multi-arch| \\
        \toprule
        \multirow{3}{*}{Our}           &  \texttt{DT}         &  39.11  &  81.40  &  77.35  &  85.90 \\
                                       &  \texttt{RF}         &  92.28  &  95.09  &  89.38  &  96.66   \\
                                       &  \cellcolor{green!40}\texttt{XGBOOST}    & \cellcolor{green!40}\textbf{96.87}   &  \cellcolor{green!40}\textbf{97.07}  &  \cellcolor{green!40}\textbf{90.80}  &  \cellcolor{green!40}\textbf{98.50}   \\
        \midrule
        \multirow{3}{*}{\shortstack[l]{Pajouh-\\based}}    
                                 &  \texttt{DT}        &  37.38            &  66.43               &  66.01             &  62.81  \\
                                 &  \texttt{RF}        &  87.68            &  80.91               &  77.18             &  88.62   \\
                                 &  \texttt{XGBOOST}   &  \textbf{88.44}   &  \textbf{82.86}      &  \textbf{78.05}    &  \textbf{90.90}  \\
        \midrule
        \multirow{3}{*}{Thaeler}  &  \texttt{DT}       &  38.15            &  40.77            &  65.48              &  72.61  \\
                                  &  \texttt{RF}       &  81.26            &  69.14            &  71.93              &  81.36   \\
                                  &  \texttt{XGBOOST}  &  \textbf{86.78}   &  \textbf{69.25}   &  \textbf{72.57}     &  \textbf{90.23}  \\
        \midrule
        MalConv & - &  \textbf{82.04}   &  \textbf{79.81}  &  \textbf{81.77}   &  \textbf{89.06}  \\
        \midrule
        Avg. Improv.  &  &  +13.07\%  & \textbf{+26.31\%}  &  +17.49\%  &  +9.39\%  \\
        \bottomrule
    \end{tabular}
    \label{tab:dr_table_final}
\end{table}

\subsection{Comparison with State-of-the-Art Detectors}
The results are presented in \autoref{tab:dr_table_final}, which shows the True Positive Rate (TPR) at 1\% False Positive Rate (FPR) for all the feature sets, models, and architectures.
They highlight several interesting takeaways.

First and foremost, our feature set consistently outperforms the others in all the considered scenarios. Specifically, considering the best-performing model for each feature set, our detector (based on XGBoost) achieves an average improvement of 13.07\%, 26.31\%, 17.49\%, and 9.39\% for the \verb|x86-64|, \verb|arm64|, \verb|fat|, and \verb|multi-arch| datasets, respectively.
On average, across all benchmarking datasets, our detector achieves a relative improvement of 16.56\% compared to the state-of-the-art detectors.
To clarify, the average improvement is computed as the mean of the relative improvements of our best detector (XGBoost) over the best-performing detector for each of the other feature sets, as $(\text{our} - \text{other}) / \text{other} \times 100$.
We used the same formula to compute the average improvement (or decrease) for all the comparisons reported below.

Notably, the DT model achieved low performance ($<$ 50\% TPR) in all scenarios,
suggesting that effective classification requires capturing complex feature interactions that can only be modeled by more advanced tree-based approaches such as Random Forest and XGBoost.
Nevertheless, the detectors were evaluated at a very low FPR (\eg, 1\%), which represents a particularly challenging operating point for malware detection systems.

Also, it is worth remarking that our detector performs especially well on the \verb|arm64| dataset (26.31\%), confirming the effectiveness of the proposed features for emerging CPU architectures.

Furthermore, the MalConv model, which uses raw byte sequences of executables as input, is the worst-performing detector in all the considered scenarios, highlighting the importance of feature engineering in macOS malware detection.

Moreover, across all the proposed detectors, the performance on the \verb|multi-arch| dataset is higher than on the other datasets.
This may be due to the fact that training on a more diverse and larger dataset enables the detectors to learn distinguishing characteristics from multiple architectures.

Finally, for completeness in Appendix~\ref{appendix:roc_curves} we also report Figures \ref{fig:roc_ml} and \ref{fig:roc_detectors} to further evaluate the performance of the target models and feature sets.
The former (\ie, \autoref{fig:roc_ml}) shows the ROC curves of the machine learning models evaluated in this work on the proposed features, namely Decision Tree (\texttt{DT}), Random Forest (\texttt{RF}), and XGBoost (\texttt{XGBOOST}).
This figure confirms that the XGBoost model achieves the best performance among the evaluated models, especially at very low false positive rates.

The latter (\ie, \autoref{fig:roc_detectors}) shows the ROC curves of the XGBoost model (the best among the evaluated models) trained on the state-of-the-art feature sets evaluated in this work, \ie, ours (\texttt{our}),
the features used in Pajouh \etal~\cite{pajouh2018intelligent} (\texttt{pajouh}), those proposed by Thaeler \etal~\cite{thaeler_macos_2024} (\texttt{thaeler}), as well as MalConv (\texttt{malconv}).
These results clearly demonstrate that the XGBoost model based on our features consistently outperforms the state-of-the-art solutions at all false positive rates, hence confirming the effectiveness of the proposed features for macOS malware detection.

\vspace{0.2cm}
\takeawaysNoTitle{
\textbf{Feature Effectiveness} --   
Thanks to the use of the new proposed features, our detector achieves an \textbf{average improvement of 16.56\%} over state-of-the-art solutions trained on generic features (\textbf{RQ.1}).
Our experiments also show that the detectors trained on the entire \texttt{multi-arch} dataset provide better results, highlighting the benefits of using larger and more diverse training samples.
}

\begin{figure*}[!t]
    \centering
    \includegraphics[width=0.91\textwidth]{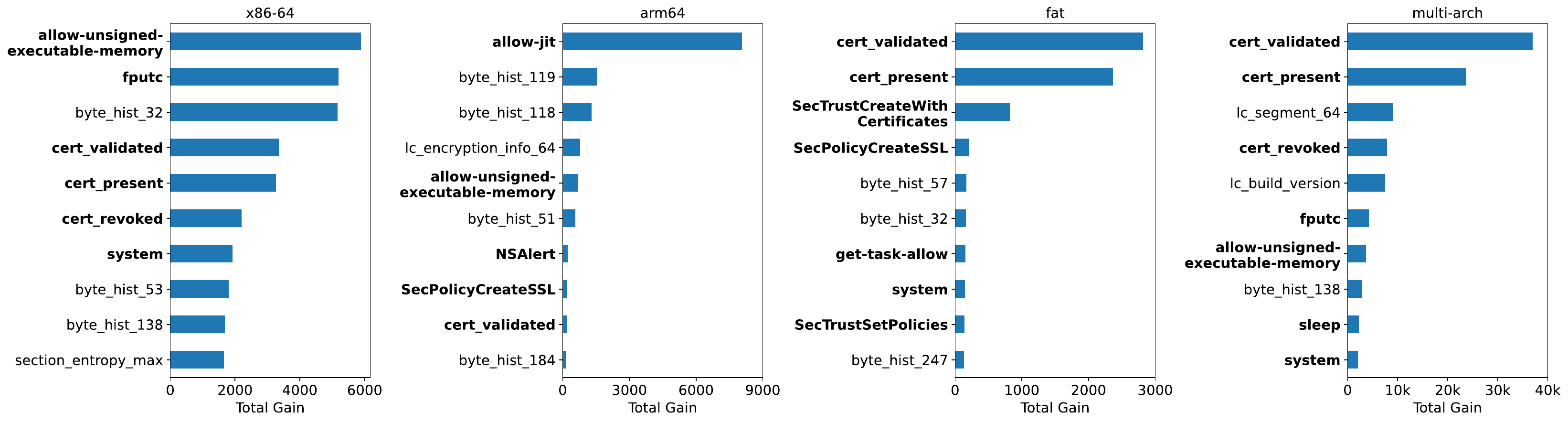}
    \caption{Feature importance based on total gain for our detector. The macOS-specific features are highlighted in bold.}
    \label{fig:feat_importance}
\end{figure*}

\subsection{Effectiveness of Domain-Specific Features}\label{subsec:feat_importance}
In this section, we conduct two studies to evaluate the effectiveness of the macOS-specific features proposed in this work.
First, we evaluate their importance, \ie, if our detector (based on the XGBoost model) effectively leverages the macOS-specific features for classification.
Then, we assess the impact of removing these features on the detection performance.

\myparagraphlb{Features Importance.}
\autoref{fig:feat_importance} shows, for each CPU architecture, the top-10 most important features based on the \emph{total gain}, a metric commonly used to identify the features that most significantly influence XGBoost predictions~\citep{xgboost_feature_importance}.
Although feature importance varies across the benchmarking datasets, in all the scenarios the macOS-specific features (emphasized in bold in the figure) play a key role in the classification task.
Notably, whether the certificate is validated (\codeinline{cert_validated}) and the presence of the certificate chain (\codeinline{cert_present}) are the most important features for the \verb|multi-arch| and \verb|fat| datasets.
This finding aligns with our preliminary analysis in \autoref{sec:dataset}, which emphasizes their relevance in distinguishing goodware from malware samples.

\begin{table}[!t]
    \centering
    \footnotesize
    \caption{TPR at 1\% FPR of our detector (\texttt{XGBOOST}) for feature importance analysis based on \emph{feature assessment}.}
    \begin{tabular}{c c c c c c}
        \toprule
        Features                  &  \verb|x86-64|       &  \verb|arm64|     &  \verb|fat|        &  \verb|multi-arch|  \\
        \midrule
        \codeinline{all} &  \textbf{96.87}       &  \textbf{97.07}    &  \textbf{90.80}     &  \textbf{98.50}  \\
        \codeinline{generic}      &  95.16                &  92.60             &  86.49              &  96.68 \\
        \codeinline{specific}     &  95.72                &  94.88             &  90.44              &  97.96 \\
        \midrule
        \codeinline{generic} vs \codeinline{all} &  -1.76\%  &  -4.60\%  &  -4.74\%  &  -1.85\%  \\
        \bottomrule
    \end{tabular}

    \label{tab:dr_analysis_features}
\end{table}

As for the \verb|x86-64| dataset, the most important feature is the presence of the \codeinline{allow-unsigned-executable-memory} entitlement, instead for \verb|arm64| the presence of the \codeinline{allow-jit} entitlement is the most important one, which has a huge impact on the model's predictions.
This is in accordance with the analysis conducted in \autoref{sec:dataset} regarding the entitlements, which shows that the presence of \codeinline{allow-jit} is a strong indicator of a sample being goodware, especially for \verb|arm64| samples.

Finally, it is worth noting that some low-level APIs, such as \codeinline{fputc} and \codeinline{system}, are among the most important features for the \verb|x86-64|, \verb|fat|, and \verb|multi-arch| datasets.
To this end, by analyzing the frequency of \codeinline{fputc}, we identified it as the most distinctive low-level API call for goodware samples, as it exhibits the largest normalized difference between the number of goodware and malware samples that utilize it.
This likely explains why the model picks up this feature as a strong indicator of goodness.
As for the \codeinline{system} API, instead, we found that it is widely used by malware (33.78\%) but very rarely by goodware (1.02\%) in our dataset.
This is somehow expected, since by analyzing some reports of macOS malware samples available on VT, we found that the \codeinline{system} API is often used to directly execute system commands.

\myparagraphlb{Feature Assessment.}
To further investigate the effectiveness of the macOS-specific features, we divided the proposed features into two subsets: \codeinline{generic} and \codeinline{specific}.
The former (\codeinline{generic}) includes only the baseline features commonly used in malware analysis, such as structural, byte-based, string-based, and packing-based features.
The latter (\codeinline{specific}) includes the macOS-specific features, \ie, certificate chain status, entitlements, persistence, and APIs.

We re-trained the XGBoost model on these two subsets by using the same experimental setup described above.
This allows us to assess the impact of removing the macOS-specific features.
The results, reported in \autoref{tab:dr_analysis_features}, show that the detector trained on all features (\codeinline{all}) performs better than those trained on only a subset of them.
When used in isolation, the \codeinline{specific} features outperform the \codeinline{generic} ones, particularly for \verb|arm64| and \verb|fat|.
However, on the full dataset (\verb|multi-arch|), removing the \codeinline{specific} features results in only a 1.85\% drop in performance (3.23\% on average across all datasets).

\vspace{0.2cm}
\takeawaysNoTitle{
\textbf{Feature importance} --
Our results show that when the macOS-specific features are available, our detector considers them very important. 
In particular, features based on the certificate status, entitlements, and system-related APIs are among the most impactful (\textbf{RQ.2}).
However, when these specific features are removed, our detector is still capable of achieving similar results by relying only on generic features.
}

\subsection{Real-world Assessment}\label{subsec:real_world_assessment}
We now provide a real-world assessment of the proposed detector by evaluating its effectiveness in the wild and its generalization capabilities over time.
To this end, we created a new \emph{fresh} dataset of 9,000 samples, including 4,500 goodware and 4,500 malware samples, extracted from the VT feed over a period of 3 months (from September to November 2024) according to the following processing steps.

\myparagraph{Samples processing.} 
We selected only samples that are \macho executables.
Samples were sorted first using the \codeinline{first_submission_ date} field of the VT report to select the most recent samples by submission date,
and then using the \codeinline{last_analysis_date} field to select the most recent samples by analysis date.
As for the malware samples, we selected only those detected by at least 5 antivirus engines, while for the goodware samples we selected only those that have zero detections.
Finally, for both goodware and malware samples, we took 4,500 samples in a stratified way, \ie proportionally to the number of samples for each architecture in the VT feed.

\myparagraph{Dataset Distribution.}
\autoref{tab:arch_distribution_fresh} shows the distribution of the samples for each CPU architecture in the \emph{fresh} dataset.
Similarly to the original dataset, the \verb|x86-64| architecture is the most represented one, with 51\% of the samples, followed by the \verb|fat| architecture with 32\% of the samples, and the \verb|arm64| architecture with 17\% of the samples.
Moreover, compared to the main dataset (see \autoref{tab:arch_distribution}), the percentages related to the malware distribution highlight an increasing number of both \verb|arm64| (16\% vs. 6\%) and \verb|fat| (15\% vs. 2\%) samples, which confirms the trend of malware authors to target these architectures~\citep{cuckoo_sentinellabs_2024}.

\begin{table}[!t]
    \centering
    \footnotesize
    \caption{Samples distribution by CPU architecture for the \emph{fresh} dataset.
    Percentages in the \emph{Total} row refer to all samples, while the others are computed relative to each label.}
    \begin{tabular}{C{1.7cm} C{1.7cm} C{1.7cm} C{1.7cm}}
          \toprule
          Label  &  \verb|x86-64|  &  \verb|arm64|  &  \verb|fat|  \\
          \toprule
          Malware    &  3,090 (68\%)  &  715 (16\%)    &  695 (15\%)  \\
          Goodware   &  1,491 (33\%)  &  801 (18\%)    &  2,208 (49\%)  \\
          \midrule
          Total    &  4,581  (51\%)   &  1,516 (17\%)  &  2,903  (32\%)   \\
          \bottomrule
    \end{tabular}

    \label{tab:arch_distribution_fresh}
\end{table}

\begin{table}[!t]
    \centering
    \footnotesize
    \caption{TPR at 1\% FPR of the detectors evaluated for the real-world assessment.
    The average improvement w.r.t. the s.o.t.a. detectors is based on the \codeinline{specific} feature set.
    }
    \begin{tabular}{c c c c c}
        \toprule
        Detector     &  \verb|x86-64|    &  \verb|arm64|     &  \verb|fat|        &  \verb|multi-arch| \\
        \midrule
        our -- \codeinline{all}          & \textbf{98.20}             & 73.47             & 97.42              & 99.50 \\
        our -- \codeinline{generic}         & 92.01             & 58.82             & 69.64              & 90.62 \\
        \textbf{our -- \codeinline{specific}}     & 97.74    & \textbf{77.55}    & \textbf{97.71}     & \textbf{99.50} \\
        \midrule
        \codeinline{generic} vs \codeinline{all}     &  -6.30\%  &  -19.94\%  &  -28.52\%  &  -8.92\%  \\
        \codeinline{specific} vs \codeinline{all}     &  -0.47\%  &  +5.55\%    &  +0.30\%    &  +0.00\%  \\
        \midrule
        Pajouh-based   &   71.98  &  54.43  &  60.86  &  68.38  \\
        Thaeler  &  70.14  &   53.77  &   57.49  &  67.72  \\
        MalConv  &  65.42  &   51.10  &   61.22  &  64.38  \\
        \midrule
        Avg. Improv. &  +37.56\%  &  +43.34\%  & +65.24\%  &  +46.21\% \\
        \bottomrule
    \end{tabular}

    \label{tab:dr_analysis_itw}
\end{table}

\begin{table}[!t]
    \centering
    \footnotesize
    \caption{F1-score at 1\% FPR of the detectors evaluated for the real-world assessment.}
    \begin{tabular}{c c c c c}
        \toprule
        Detector                &  \verb|x86-64|    &  \verb|arm64|     &  \verb|fat|        &  \verb|multi-arch| \\
        \midrule
        our -- \codeinline{all}        & 94.92             & 75.91             & 96.41              & 96.74 \\
        our -- \codeinline{generic}       & 92.38             & 64.93             & 80.00              & 90.15 \\
        \textbf{our -- \codeinline{specific}}   & \textbf{97.30}    & \textbf{80.43}    & \textbf{96.52}     & \textbf{98.50} \\
        \midrule
        \codeinline{generic} vs \codeinline{all}     &  -2.68\%  &  -14.45\%  &  -17.02\%  &  -6.81\%  \\
        \codeinline{specific} vs \codeinline{all}  &  +2.51\%   &  +5.95\%    &  +0.11\%    &  +1.82\%  \\
        \midrule
        Pajouh-based   &  81.83  &  60.12   &   64.10   &   75.21  \\
        Thaeler        &  79.52  &  56.75   &   62.84   &   72.85  \\
        MalConv        &  75.27  &  54.48   &   66.80   &   70.55 \\
        \midrule
        Avg. Improv. &  +20.62\%  &  +37.75\%  & +52.08\%  &  +33.08\% \\
        \bottomrule
    \end{tabular}

    \label{tab:f1_analysis_itw}
\end{table}

\myparagraph{Evaluation.}
We used the \emph{fresh} dataset to evaluate the generalization performance of our detector in the wild.
For completeness, we also evaluate the detectors based on the features proposed by~\cite{pajouh2018intelligent} and~\cite{thaeler_macos_2024}, as well as MalConv~\citep{malconv2}.
Moreover, as already done in the previous experiments, we evaluated different combinations of the proposed features, namely \codeinline{all}, \codeinline{generic}, and \codeinline{specific}.

All the detectors are trained on the main dataset.
Furthermore, the classification threshold was computed on the test set of the main dataset and then used to evaluate the detectors on the \emph{fresh} dataset.
The results are reported in \autoref{tab:dr_analysis_itw} and \autoref{tab:f1_analysis_itw}, which show the detection rate (\ie, TPR) and F1-score at 1\% FPR, respectively.

The tables highlight several interesting takeaways.

First, unlike the \emph{feature assessment} experiment, in this case, removing the \texttt{specific} features leads to a significant drop in detection performance (15.92\% on average).
This is particularly evident for the \verb|arm64| and \verb|fat| datasets, where the relative performance drop is 19.94\% and 28.52\%, respectively.
This confirms that the \texttt{specific} features generalize significantly better than the \texttt{generic} ones on novel malware samples.

Second, compared to the \codeinline{all} feature set, the \codeinline{specific} one achieves better performance in terms of F1-score across all the benchmarking datasets (2.60\% on average),
as well as a slight improvement in the TPR (1.34\% on average).
The higher improvement in terms of F1-score is because the \codeinline{specific} features generates less false positives, hence increasing the precision of the model.
This can be explained by the fact that the additional \codeinline{generic} features in the \codeinline{all} feature set lead to overfitting and, consequently, a performance drop due to the curse of dimensionality~\citep{bishop_ml}.

Third, the detector based on our (\codeinline{specific}) features
consistently outperforms the state-of-the-art solutions (\ie, the detectors based on the features proposed by Pajouh \etal~\cite{pajouh2018intelligent}, Thaeler \etal~\cite{thaeler_macos_2024}, and MalConv~\cite{malconv2}) across all the considered scenarios, with a
stunning average improvement of 50.03\% and 37.34\% in terms of TPR and F1-score,
respectively.

Finally, despite the dominance of the \verb|x86-64| architecture, our detector is still able to effectively detect malware samples belonging to the other architectures too (see \verb|multi-arch| scenario).
However, we also noticed that all the detectors experience a performance drop when trained only on the \verb|arm64| dataset.
A possible explanation is that such drop is due to (i) the difference in the data distribution between the main and real-world dataset (6\% vs 16\% underlining the increasing trend of arm64 samples) and the (ii) data drift (arm64 samples in the real-world dataset differ from those in the main one).
As for this second point, we analyzed the distribution of the most important features for arm64 and found that while in the main dataset the presence of \codeinline{allow-jit} entitlement (most important feature) effectively discriminates goodware from malware, this is no longer the case in the real-world dataset, where only 5\% of goodware use this feature, while 0.05\% of malware samples have it.
Because of this, the model can no longer rely on this feature, leading to a decrease in detection performance.
To mitigate this issue, we could leverage continual learning techniques~\cite{Wang2024ContinualLearningSurvey} to continuously adapt the model to the evolving data distribution.

To further evaluate the generalization robustness of our detector, we finally performed three additional experiments.
First, we assessed its performance on goodware samples whose certificate chain has not been validated by Apple and obtained a FPR of 2.4\%, showing solid performance even when signing information is unavailable.
Second, we evaluated the impact of packing on detection performance and we found that our detector achieved 97.5\% accuracy on packed malware, a scenario that remains relatively uncommon in our real-world dataset (only $\sim$7\% of all malware).
Third, we evaluated our detector on all the signed malware included in the real-world dataset and obtained a detection rate of 96.25\%, confirming its effectiveness even against signed malware.

\vspace{0.2cm}
\takeawaysNoTitle{
\textbf{Feature Robustness --}
While on the main dataset, the removal of the macOS-specific features leads to a limited decrease (3.23\%) in detection performance,
on newly-collected samples the performance of the generic features significantly drops (15.92\%). 
This is due to the nature of the generic features (such as N-grams), which provide a powerful way to build ``signatures'' of existing malware,
but tend to poorly generalize to new samples.

On the other hand, the macOS-specific features are more semantically-rich and thus can maintain their efficacy even in presence of new variants (\textbf{RQ.3}).
This is evident when comparing our detector with state-of-the-art solutions based on generic features: when tested on novel samples, 
our detector consistently outperforms the state-of-the-art with a remarkable average improvement of 50.03\% in terms of TPR (\textbf{RQ.4}).
}

%% file: src/6-related.tex
\section{Related Work}\label{sec:related}
The detection of macOS malware through machine learning has become an area of growing interest, yet existing research exhibits notable limitations in dataset scale, adopted features, and methodological approaches (see \autoref{tab:related_work} in \autoref{appendix:related_work}).

In one of the earliest works, Pajouh \etal~\cite{pajouh2018intelligent} leveraged a dataset of 602 samples, including 152 malware and 450 goodware, to experiment with several machine learning models such as Naive Bayes, Support Vector Machine (SVM), and Decision Tree (DT).
They extracted only structural features from the Mach-O file format, such as the number of load commands, segments, symbols, and imported libraries.
Due to the limited dataset size, they employed the Synthetic Minority Over-sampling Technique (SMOTE)~\citep{smote}, which improved accuracy but also increased the false positive rate.
The same dataset was reused by subsequent works~\citep{Sahoo2022, Chen2022MLOSXMalware, Gharghasheh2022}, which further experimented with additional models such as Logistic Regression (LR) and Random Forest (RF).

In the most recent work to date, Thaeler \etal~\cite{thaeler_macos_2024} collected a new dataset comprising 852 malware and 32,333 goodware samples,
and employed a combination of structural (e.g., entropy, file size, number of load commands) and string-based (e.g., presence of suspicious strings) features to train a variety of machine learning models, including SVM, DT, and RF.

Prior studies share common limitations. They rely on small, proprietary datasets, which limits the generalizability of their findings.
Their solutions are also limited to a narrow set of structural and string-based features, overlooking critical macOS-specific characteristics such as entitlements, persistence techniques, and embedded certificate status, which, as demonstrated in this work, are essential for achieving both high detection performance and human interpretability.
Moreover, unlike this study, they do not rigorously assess feature importance, which is key to understanding the detector's decisions, nor do they evaluate how performance evolves over time by testing on a fresh dataset collected after model training.

%% file: src/7-conclusions.tex
\section{Conclusions}\label{sec:conclusions}

In the vast majority of cases, attackers must rely on OS-specific services and functionalities 
to carry out harmful actions, making such features a valuable indicator for effective malware classification. 
This holds true for macOS as well, where malicious binaries often leverage low-level APIs to interact with system frameworks and achieve unauthorized persistence.
Identifying these essential traits is fundamental to improve detection accuracy, as they directly reflect the intent and capabilities of the malware.
In this work, we focused on extracting and analyzing macOS-specific features, being the first to conduct a dedicated study in this area. 
We presented a novel machine learning-based macOS malware detector leveraging static features derived from the Mach-O file format and macOS domain knowledge, such as embedded certificates, entitlements, and persistence techniques.
Trained on a large-scale dataset of 41,129 samples, including 11,413 goodware and 29,716 malware, our detector achieves state-of-the-art detection capabilities (98.50\% TPR at 1\% FPR), outperforming existing approaches by 16.56\%.
Our detailed feature importance analysis highlights the key role of macOS-specific features, while real-world evaluation on a \emph{fresh} dataset of temporally newer samples confirms its stunning detection performance (99.50\%),
corresponding to a 50.03\% improvement over the state-of-the-art, and demonstrates the outstanding generalization capabilities of macOS-specific features (15.92\% drop in detection rate if excluded) compared to generic ones.

Overall, we strongly believe that our work represents a significant step forward in macOS malware detection, providing the first concrete example of how domain knowledge about macOS binaries can be harnessed to build a machine learning-based detector with state-of-the-art detection capabilities.

As future work, since malware authors are continuously evolving their techniques to evade detection, we plan to investigate the adversarial robustness of our detector by exploring both current~\citep{demetrio2021adversarialwin} and novel attacks specifically targeting the proposed macOS features.

\section{Acknowledgments}
This work has benefited from a government grant managed by the National Research Agency under France 2030 with reference ``ANR-22-PECY-0007''.
\newpage

%% file: src/appendix.tex
\section*{Appendix}
\section{Family Classification}\label{sec:family_classification}
We further evaluated the effectiveness of our detector (\ie, the XGBoost model based on the proposed features) for family classification.
To this end, we built a multi-class dataset consisting of all the goodware samples (11,413) from the original dataset, while for the malware samples, we considered the eight most common malware families in our dataset 
(see \autoref{tab:family_distribution}), \ie, \verb|bundlore|, \verb|adload|, \verb|pirrit|, \verb|jailbreak|, \verb|evilquest|, \verb|lador|, \verb|genieo|, and \verb|stealer|, for a total of 21,679 malware samples.
Hence, in total, the multi-class dataset consists of 33,097 samples and 9 classes (8 malware families and 1 goodware class).
To split the dataset into training and test sets, we also adopted a stratified approach to ensure that the distribution of the classes is the same in both the training and test sets.
The experimental results are reported in \autoref{tab:dr_table_families}, which shows the detection rate (\ie, the TPR) for each category, computed using the one-vs-all strategy,
\ie, we turned the multi-class problem into multiple binary classification problems, where we trained a binary classifier for each target class and then computed the TPR at 1\% FPR for each class.
The results demonstrate that the proposed detector can also effectively distinguish between the main macOS malware families, achieving an average detection rate of 99.53\%, 
with the best-performing families being \verb|evilquest| and \verb|lador|, which achieve a perfect detection rate of 100.00\%, while the worst-performing family is \verb|stealer|, which achieves a detection rate of 98.50\%, possibly due to the low number of samples available in the dataset.

\section{Hyper-parameters}\label{appendix:hyperparameters}
\autoref{tab:hyperparams} shows the hyper-parameters of the tree-based machine learning models evaluated in this work,
namely Decision Tree (\texttt{DT}), Random Forest (\texttt{RF}), and XGBoost (\texttt{XGBOOST}).
The hyper-parameters are tuned using a grid search approach, where we evaluate all the possible combinations of the hyper-parameters in the specified ranges.

\section{ROC Curves}\label{appendix:roc_curves}
\begin{figure*}[!th]
  \centering
  \includegraphics[width=0.95\textwidth]{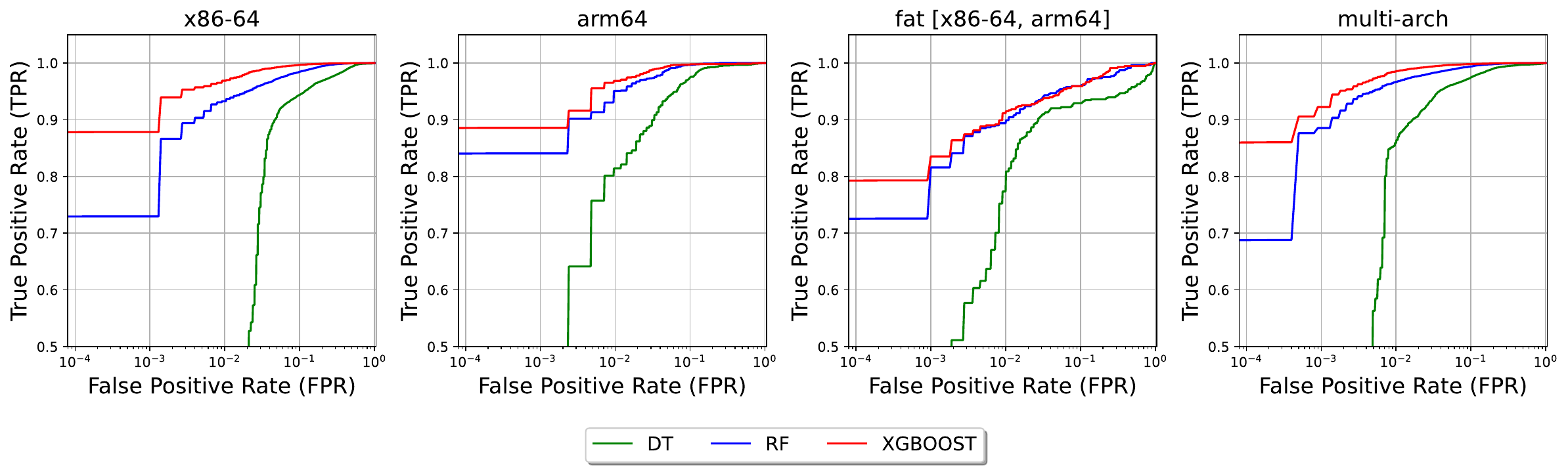}
  \caption{ROC curves of the machine learning models trained on the features proposed in this work, namely Decision Tree (\texttt{DT}), Random Forest (\texttt{RF}), and XGBoost (\texttt{XGBOOST}).
  }
  \label{fig:roc_ml}
\end{figure*}

\begin{figure*}[!th]
  \centering
  \includegraphics[width=0.90\textwidth]{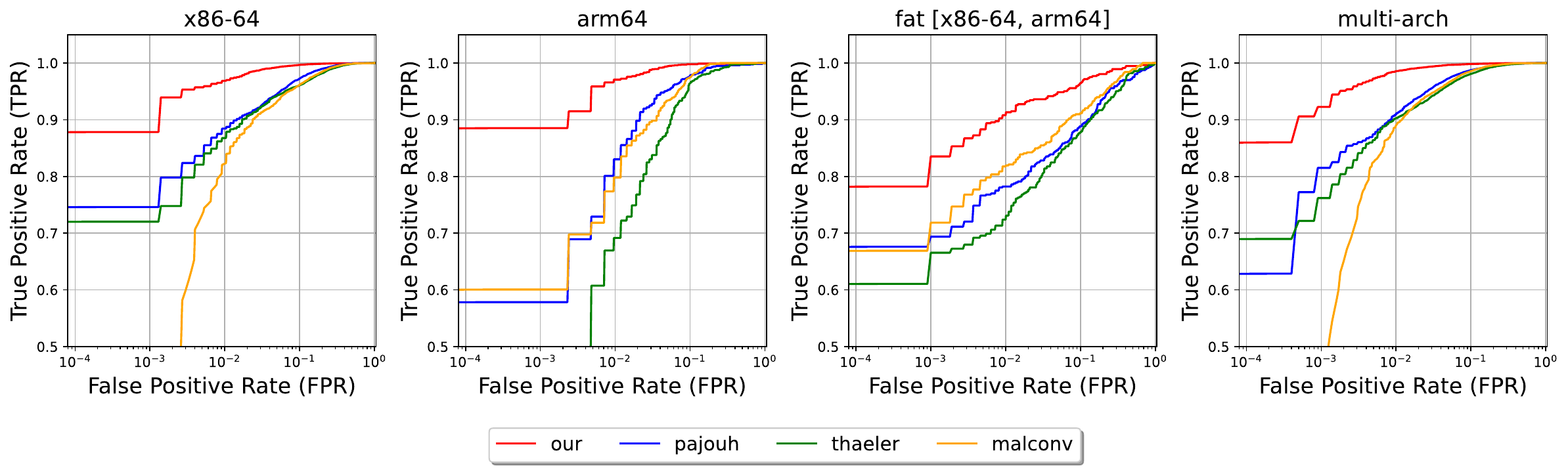}
  \caption{
  ROC curves of the XGBoost model trained on the state-of-the-art features sets, namely ours (\texttt{our}), the features used in Pajouh \etal~\cite{pajouh2018intelligent} (\texttt{pajouh}), those proposed in Thaeler \etal~\cite{thaeler_macos_2024} (\texttt{thaeler}),
  as well as MalConv (\texttt{malconv})
  }
  \label{fig:roc_detectors}
\end{figure*}

To further support the experimental results presented in \autoref{sec:exp_results}, we provide additional results in \autoref{fig:roc_ml} and \autoref{fig:roc_detectors}.
The former (\autoref{fig:roc_ml}) reports the ROC curves of the machine learning models evaluated in this work on the proposed features, namely Decision Tree (\texttt{DT}), Random Forest (\texttt{RF}), and XGBoost (\texttt{XGBOOST}).
It shows that the XGBoost model achieves the best performance among the evaluated models, especially at very low false positive rates.

The latter (\autoref{fig:roc_detectors}) shows the ROC curves of the XGBoost model (the best among the evaluated models) trained on the state-of-the-art feature sets evaluated in this work, \ie, ours (\texttt{our}), the features used in Pajouh \etal~\cite{pajouh2018intelligent} (\texttt{pajouh}), those proposed in Thaeler \etal~\cite{thaeler_macos_2024} (\texttt{thaeler}),
as well as MalConv (\texttt{malconv}).

These results clearly demonstrate that the XGBoost model based on our features consistently outperforms the state-of-the-art solutions at all false positive rates, hence confirming the effectiveness of the proposed features for macOS malware detection.

\begin{table}[!t]
  \centering
  \footnotesize
  \caption{TPR at 1\% FPR of our detector trained for family-classification.}
  \begin{adjustbox}{max width=\columnwidth}
  \begin{tabular}{c c c c c}
    \toprule
    \verb|bundlore|  &  \verb|adload|  &  \verb|pirrit|  &  \verb|jailbreak|  &  \verb|evilquest|  \\
    \midrule
    99.95            &  99.09          &  99.93          &  99.25             &  100.00 \\
    \toprule
    \verb|lador|  &  \verb|genieo|  &  \verb|stealer| &  \verb|goodware| \\
    \midrule
    100.00        &  99.52          &  98.52          &   99.96         \\
    \toprule
  \end{tabular}
  \end{adjustbox}
  \label{tab:dr_table_families}
\end{table}

\begin{table}[!t]
  \centering
  \footnotesize
  \caption{Hyper-parameters of the machine learning models evaluated in this work.}
  \begin{tabular}{c c c}
      \toprule
      \textbf{Model} & \textbf{Feature} & \textbf{Range} \\
      \toprule
      \multirow{5}{*}{\shortstack{Decision \\ Tree}}
      & \texttt{max\_depth} & [2, 10] \\
      & \texttt{max\_features} & [\texttt{sqrt}, \texttt{log2}, \texttt{None}] \\
      & \texttt{criterion} & [\texttt{gini}, \texttt{entropy}, \texttt{log\_loss}] \\
      & \texttt{min\_samples\_leaf} & [4, 8] \\
      & \texttt{min\_samples\_split} & [2, 10] \\
      \midrule
      \multirow{6}{*}{\shortstack{Random \\ Forest}}
      & \texttt{n\_estimators} & [5, 100] \\
      & \texttt{max\_depth} & [2, 10] \\
      & \texttt{max\_features} & [\texttt{sqrt}, \texttt{log2}, \texttt{None}] \\
      & \texttt{criterion} & [\texttt{gini}, \texttt{entropy}, \texttt{log\_loss}] \\
      & \texttt{min\_samples\_leaf} & [4, 8] \\
      & \texttt{min\_samples\_split} & [2, 10] \\
      \midrule
      \multirow{5}{*}{\shortstack{XGBoost}}
      & \texttt{n\_estimators} & [5, 100] \\
      & \texttt{max\_depth} & [2, 10] \\
      & \texttt{eta} & [1e-5, 1e-1] \\
      & \texttt{min\_child\_weight} & [8, 16] \\
      & \texttt{colsample\_bytree} & [0.4, 1.0] \\
      \bottomrule
  \end{tabular}

  \label{tab:hyperparams}
\end{table}

\begin{table}[!t]
    \centering
    \footnotesize
    \caption{Related work on machine learning for macOS malware detection. Features: S: structural, T: string-based, V: various (also macOS-specific)}
    \begin{adjustbox}{width=\columnwidth}
    \begin{tabular}{c c c c c c}
        \toprule
        \textbf{Work} & \textbf{Year} & \multicolumn{3}{c}{\textbf{Dataset}} & \textbf{Features} \\
        &  &  Malware  &  Goodware  &  Available  &  \\
        \toprule
        Pajouh \etal~\cite{pajouh2018intelligent} & 2018 & 152 & 450 & \xmark & \makecell{S,T} \\
        \midrule
        Sahoo \etal~\cite{Sahoo2022} & 2022 & 152 & 450 &  \xmark  &  \makecell{S,T} \\
        \midrule
        Chen \etal~\cite{Chen2022MLOSXMalware} & 2022 & 152 & 450 & \xmark & \makecell{S,T} \\
        \midrule
        Gharghasheh \etal~\cite{Gharghasheh2022} & 2022 & 152 & 450 & \xmark & \makecell{S,T} \\
        \midrule
        Thaeler \etal~\cite{thaeler_macos_2024}  & 2023 & 852 & 32,333 & \xmark & \makecell{S,T} \\
        \midrule
        This work & 2025 &  29,716  &  11,413  & \cmark & \makecell{V} \\
        \bottomrule
    \end{tabular}
    \label{tab:related_work}
    \end{adjustbox}
\end{table}

\section{Related Work}\label{appendix:related_work}
To further contextualize our work and complement the discussion in \autoref{sec:related}, we provide
\autoref{tab:related_work} that summarizes the main characteristics of the existing works on machine learning for macOS malware detection.
In particular, it reports the number of malware and goodware samples in the dataset used by each work, whether the dataset is publicly available, and the type of features adopted for training the models.